\documentclass[twocolumn,useAMS,usenatbib]{aastex61}
\usepackage{epsfig, bm, times, aas_macros}
\usepackage[english]{babel}
\usepackage{amsmath}
\usepackage{amssymb}
\usepackage[retainorgcmds]{IEEEtrantools}
\usepackage{graphicx}
\usepackage{ifthen}
\usepackage{soul}

\newcommand{\mdotgen}{\dot m}
\newcommand{\mdot}{\mdotgen_{\rm{p}}}
\newcommand{\Mdot}{\dot M}
\newcommand{\mdotedd}{\mdotgen_{\rm {edd}}}
\newcommand{\locmdot}{\mdotgen(\theta, \nu)}
\newcommand{\g}{av}
\newcommand{\mdotg}{\mdotgen_{\rm{\g}}}

\newcommand{\mdott}{\Mdot_{\rm{tot}}}
\newcommand{\mdottmax}{\Mdot_{\rm{tot,\, max}}}
\newcommand{\Rs}{R_\star}

\newcommand{\Ms}{M_\star}
\newcommand{\nuk}{\nu_{\rm{k}}}
\newcommand{\gefbase}{g_{\rm{eff}}}
\newcommand{\gefbaseE}[1]{g_{\rm{eff, #1}}}
\newcommand{\gep}{\gefbaseE{p}}

\newcommand{\gef}[1]{\gefbase^{#1}}
\newcommand{\gefd}[1]{\gefbase(\theta,\nu)^{#1}}
\newcommand{\gfe}{\gef{}}
\newcommand{\gfed}{\gefd{}}
\newcommand{\xbase}{\bar{g}}
\newcommand{\xbased}{\xbase(\theta,\nu)}
\newcommand{\x}[1]{\xbase^{#1}}
\newcommand{\xd}[1]{\xbased^{#1}}
\newcommand{\alp}{\alpha}
\newcommand{\alpt}{A}
\newcommand{\bet}{\beta}
\newcommand{\bett}{B}
\newcommand{\gam}{\gamma}
\newcommand{\gamt}{\Gamma}
\newcommand{\low}{\rm{l}}
\newcommand{\hig}{\rm{h}}
\newcommand{\gamtl}{\gamt_{\low}}
\newcommand{\gamth}{\gamt_{\hig}}
\newcommand{\gaml}{\gam_{\low}}
\newcommand{\gamh}{\gam_{\hig}}
\newcommand{\mlt}{\locmdot_{\low}}
\newcommand{\mht}{\locmdot_{\hig}}
\newcommand{\ml}{\mdotgen_{\low}}
\newcommand{\mh}{\mdotgen_{\hig}}
\newcommand{\R}{\mathcal{R}}
\newcommand{\Rb}{\bar \R}
\newcommand{\ii}{{\rm{i}}}
\newcommand{\iip}{{\rm{i+1}}}
\newcommand{\gami}{\gam_{\ii}}
\newcommand{\gamip}{\gam_{\iip}}
\newcommand{\mi}{\mdotgen_{\ii}}
\newcommand{\mip}{\mdotgen_{\iip}}
\newcommand{\alpi}{\alpha_\ii}
\newcommand{\beti}{\beta_\ii}
\newcommand{\Ri}{\R_\ii}
\newcommand{\Rip}{\R_{\iip}}
\newcommand{\Rbi}{\bar \Ri}
\newcommand{\Rii}{\R_{\ii,\ii}}
\newcommand{\Riip}{\R_{\ii,\iip}}
\newcommand{\del}{\delta}
\newcommand{\delii}{\del_{\ii,\ii}}
\newcommand{\deliip}{\del_{\ii,\iip}}
\newcommand{\delgami}{\Delta\gami}
\newcommand{\rhoi}{\mu_\ii}
\newcommand{\equi}{\star}
\newcommand{\thesi}{\theta^\equi_\ii}
\newcommand{\xsireal}{{\xbase^\equi_\ii}}
\newcommand{\xmin}{{\xbase_{\rm{min}}}}
\newcommand{\xmax}{{\xbase_{\rm{max}}}}

\newcommand{\xbi}{\xbase_{\ii,\ii}}
\newcommand{\xbip}{\xbase_{\ii,\iip}}
\newcommand{\xbl}{\xbase_{\ii,\low}}
\newcommand{\xbh}{\xbase_{\ii,\hig}}

\newcommand{\anda}{\;\&\;}

\newcommand{\ep}{\epsilon}
\newcommand{\eps}[1]{\ep^{#1}}
\newcommand{\gamtn}[1]{\gamt_{#1}}
\newcommand{\gamn}[1]{\gam_{#1}}
\newcommand{\alptn}[1]{\alpt_{#1}}
\newcommand{\alpn}[1]{\alp_{#1}}
\newcommand{\bettn}[1]{\bett_{#1}}
\newcommand{\betn}[1]{\bet_{#1}}
\newcommand{\delgamn}[1]{\Delta\gam_{#1}}
\newcommand{\delnn}[2]{\del_{#1,#2}}

\newcommand{\epsstar}{{\ep{}}^{\star}}

\newcommand{\epsstari}{\epsstar_\ii}
\newcommand{\xsi}{\epsstari}

\newcommand{\mix}{\rm{m}}

\newcommand{\xbasem}{\bar{g}_{\mix}}

\newcommand{\xm}[1]{\xbasem^{#1}}

\newcommand{\themsi}{\theta^\equi_{\mix, \ii}}
\newcommand{\epm}{\epsilon_{\mix}}
\newcommand{\num}{\nu_{\rm{min}}}
\newcommand{\betm}{\bet_{\mix}}
\newcommand{\gamml}{\gam_{\mix, \low}}
\newcommand{\gammh}{\gam_{\mix, \hig}}
\newcommand{\gammg}[1]{\gam_{\mix, #1}}

\newcommand{\lt}{<}

\newcommand{\zcn}{Z_{\rm{CNO}}}
\newcommand{\yig}{y_{\rm{ign}}}
\newcommand{\yi}[1]{y_{#1,\,\rm{ign}}}
\newcommand{\ms}{M$_\odot$}

\newcommand{\Hz}{\;\textrm{Hz}}
\newcommand{\gcm}{\;\textrm{g cm}^{-2}}
\newcommand{\gscm}{\;\textrm{g s$^{-1}$ cm}^{-2}}

\newcommand{\spsp}{\;\;}

\newcommand{\trextra}[1]{\;\textrm{#1}\;}

\newcommand{\rd}{\rm{d}}

\newcommand{\seclab}{Section}

\newcommand{\applab}{Appendix}

\newcommand{\figlab}{Fig}
\newcommand{\equlab}{Eq}
\newcommand{\secref}[1]{\seclab{} \ref{#1}}
\newcommand{\secrefs}[1]{\seclab{s} \ref{#1}}

\newcommand{\appref}[1]{\applab{} \ref{#1}}

\newcommand{\figref}[1]{\figlab{.} \ref{#1}}
\newcommand{\figrefs}[1]{\figlab{s.} \ref{#1}}
\newcommand{\eqr}[1]{\equlab{.} \eqref{#1}}
\newcommand{\eqrs}[1]{\equlab{s.} \eqref{#1}}
\newcommand{\alienequlab}{eq}
\newcommand{\alienequ}[1]{\alienequlab{.} (#1)}

\newcommand{\tent}[1]{\times 10^{#1}}
\newcommand{\citeple}[1]{\citeauthor{#1}, \citeyear{#1}}

\newcommand{\uplp}[1]{\textbf{#1}}

\newcommand{\segu}{$\overline{\rm{AD}}$}
\newcommand{\segd}{$\overline{\rm{BC}}$}
\newcommand{\segt}{$\overline{\rm{AB}}$}
\newcommand{\segq}{$\overline{\rm{DC}}$}
\newcommand{\Ap}{$\rm{A}$}
\newcommand{\Bp}{$\rm{B}$}
\newcommand{\Cp}{$\rm{C}$}
\newcommand{\Dp}{$\rm{D}$}
\newcommand{\track}{path}

\newcommand{\captionoffigure}[1]{%
  \caption{#1}
}

\newcommand{\twofiginarow}[3]{
  \hspace{\stretch{1}}
  \includegraphics[width=#3\textwidth]{#1}
  \hspace{\stretch{1}}
  \includegraphics[width=#3\textwidth]{#2}
  \hspace{\stretch{1}}$\phantom{a}$
}

\newcommand{\afig}[3]{
  \begin{figure*}
    \centering
    \includegraphics[width=0.9\textwidth]{#1}
    \captionoffigure{#3}
    \label{#2}
  \end{figure*}
}

\newcommand{\onepafig}[6]{
                 {
                   \afig{#1}{#2}{#3}
                   \afig{#4}{#5}{#6}
                 }
}

\newcommand{\figzerapp}{
  \onepafig{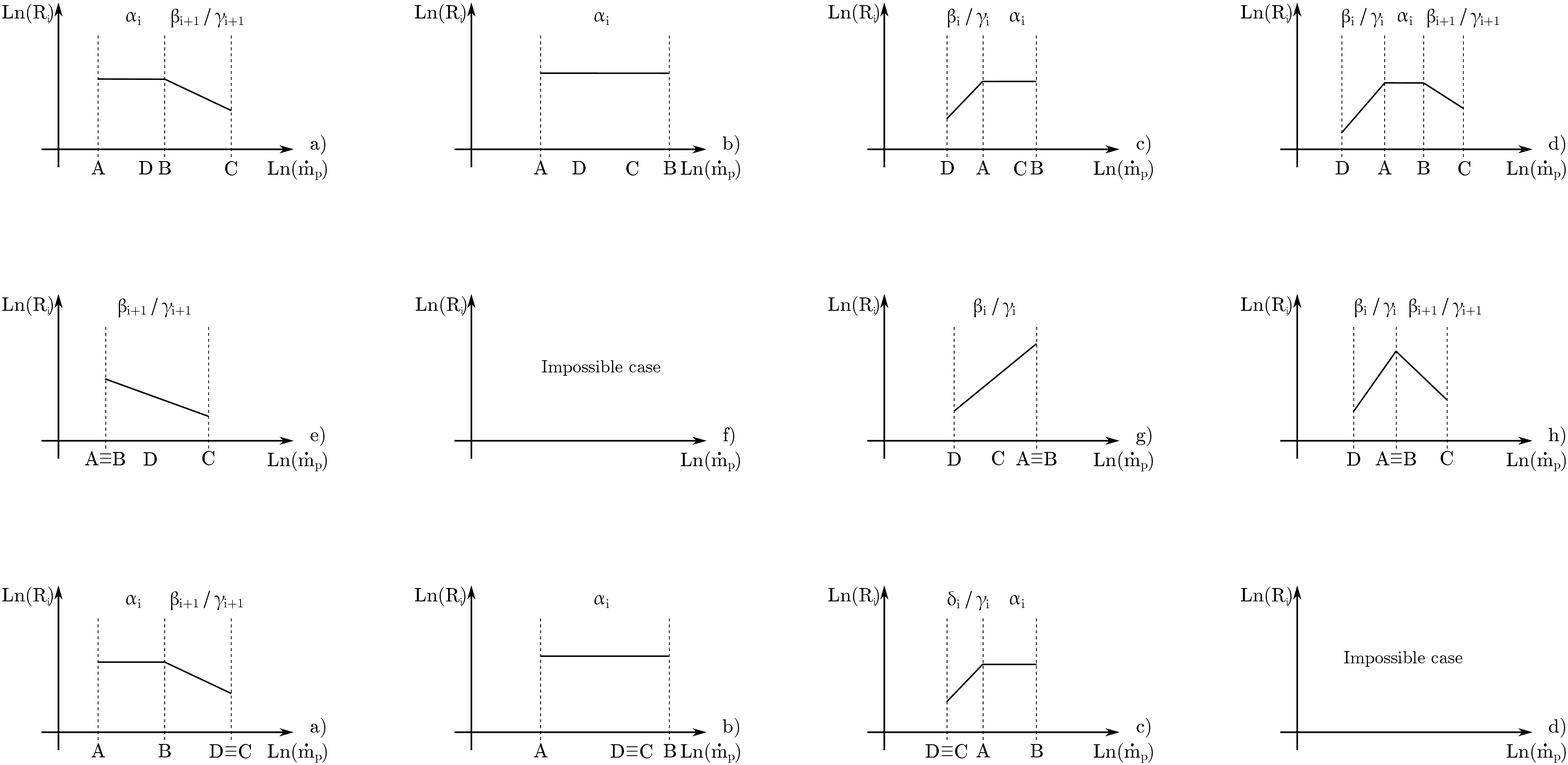}{fig:zerRBneg}%
           {Same as \figref{fig:RBneg}, $\beti < 0$, but for cases
             when $\alpi=0$. \uplp{Top and middle row}: these burst
             rate trends correspond to the blue solid \track{s} in
             \figref{fig:TOS}. Both the signs of $\deliip/\gamip =
             \beti/\gamip$ and $\delii/\gami = \beti/\gami$ are
             known. \uplp{Bottom row}: burst rate corresponding to the
             blue solid paths of \figref{fig:lessone}, where
             $\rhoi<1$.}%
           {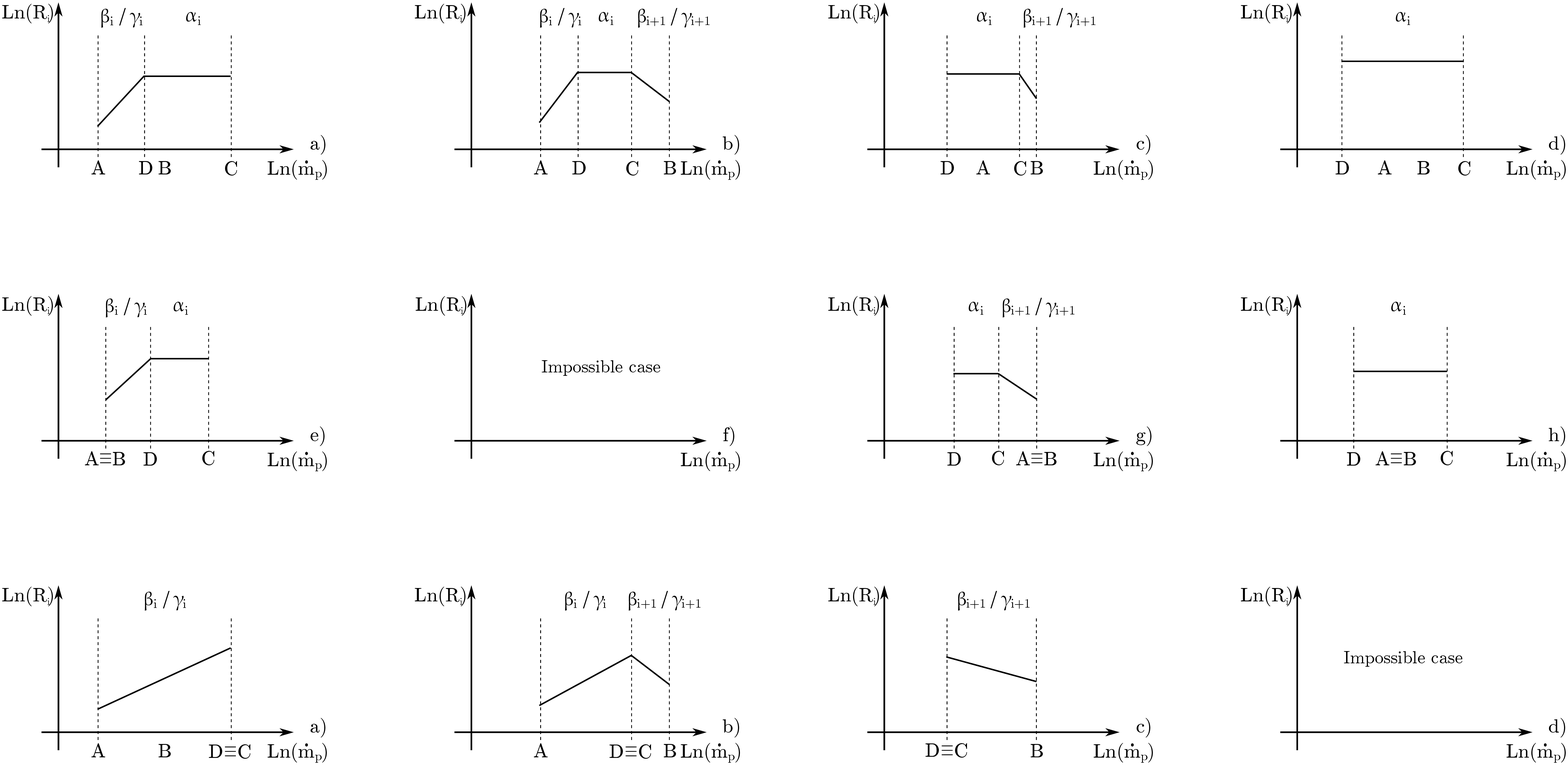}{fig:zerRBpos}%
           {Same as \figref{fig:RBpos}, $\beti > 0$, but for cases
             when $\alpi=0$. \uplp{Top and middle row}: these burst
             rate trends correspond to the red dashed \track{s} of
             \figref{fig:TOS}. Both the signs of $\deliip/\gamip =
             \beti/\gamip$ and $\delii/\gami = \beti/\gami$ are
             known. \uplp{Bottom row}: burst rate corresponding to the
             red dashed paths of \figref{fig:lessone}, where
             $\rhoi<1$.}
}

\newcommand{\fignegapp}{
  \onepafig{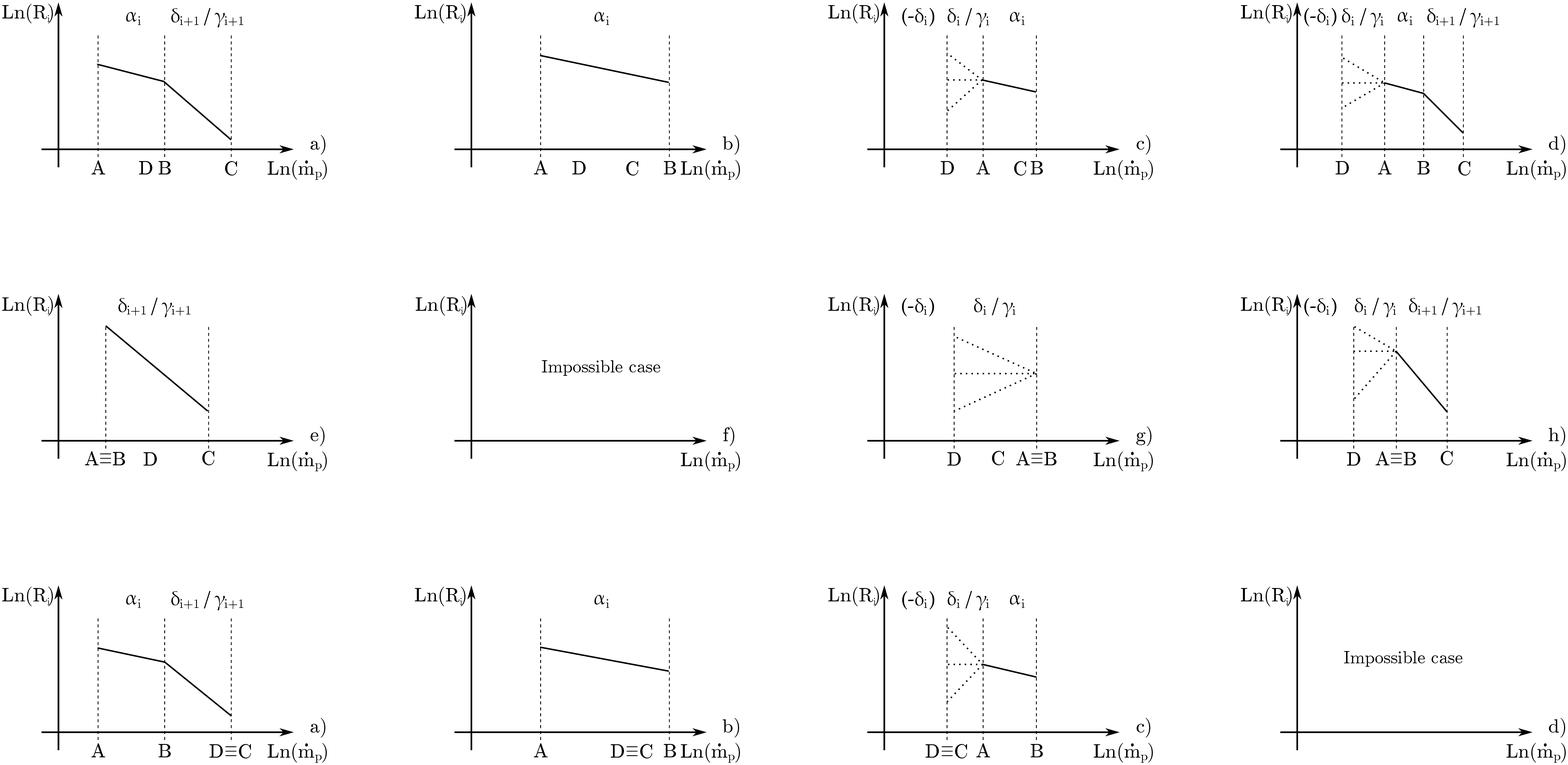}{fig:negRBneg}%
           {Same as \figref{fig:RBneg}, $\beti < 0$, but for cases
             when $\alpi<0$. \uplp{Top and middle row}: these burst
             rate trends correspond to the blue solid \track{s} in
             \figref{fig:TOS}. In this case the sign of
             $\deliip/\gamip$ is known when we need it and
             $\deliip/\gamip<\alpi<0$. The sign of $\delii/\gami$ is
             not known when we need it and it depends on the sign of
             $-\delii$. \uplp{Bottom row}: burst rate corresponding to
             the blue solid paths of \figref{fig:lessone}, where
             $\rhoi<1$.}%
           {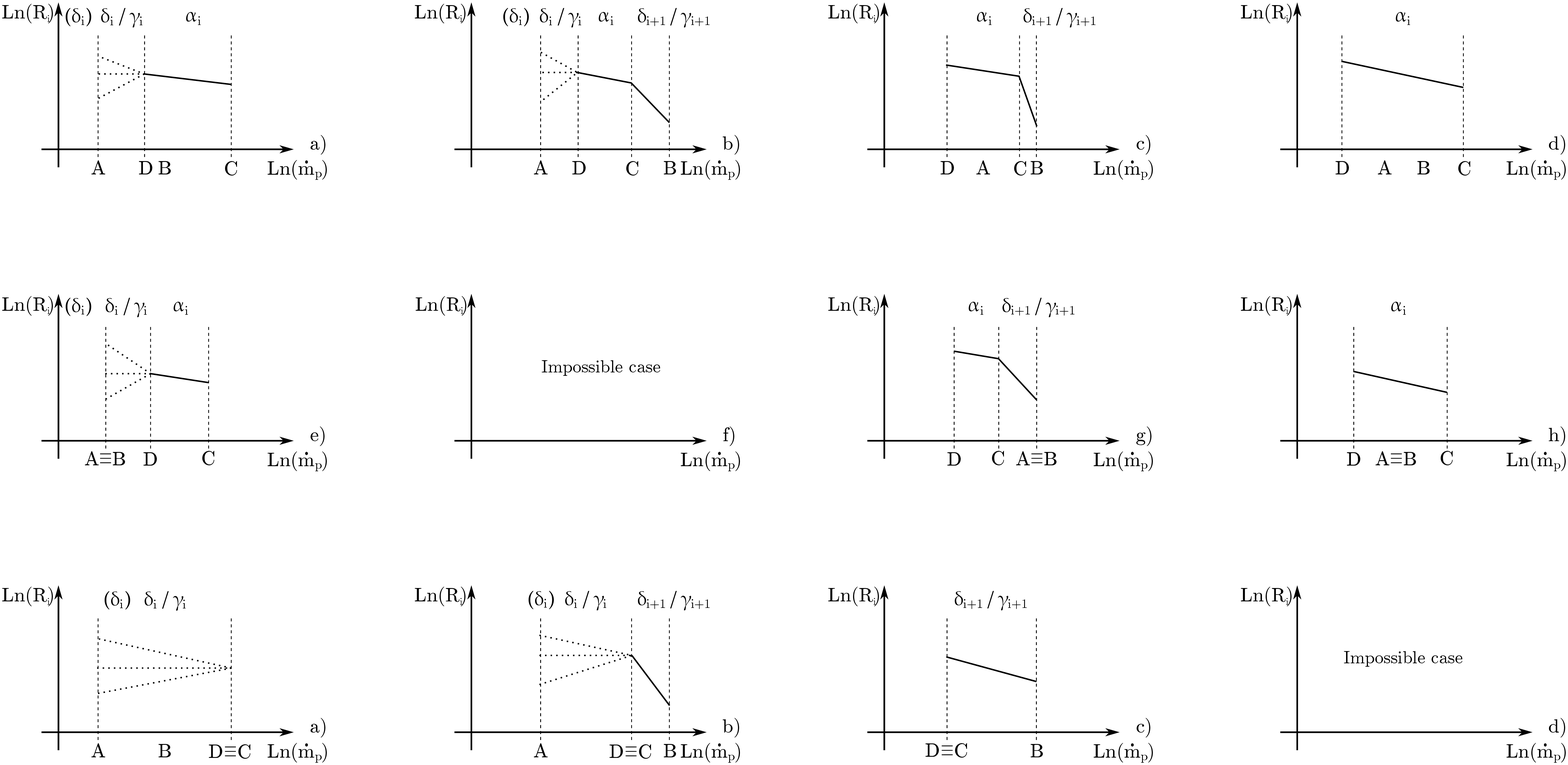}{fig:negRBpos}%
           {Same as \figref{fig:RBpos}, $\beti > 0$, but for cases
             when $\alpi<0$. \uplp{Top and middle row}: these burst
             rate trends correspond to the red dashed \track{s} of
             \figref{fig:TOS}. It is known that
             $\deliip/\gamip<\alpi<0$ in the cases where we need
             it. While the sign of $\delii/\gami$ is not known when we
             need it and it depends on the sign of
             $\delii$. \uplp{Bottom row}: burst rate corresponding to
             the red dashed paths of \figref{fig:lessone}, where
             $\rhoi<1$.}
}

\newcommand{\figlessone}{
  \begin{figure*}
    \centering
    \includegraphics[width=0.95\textwidth]{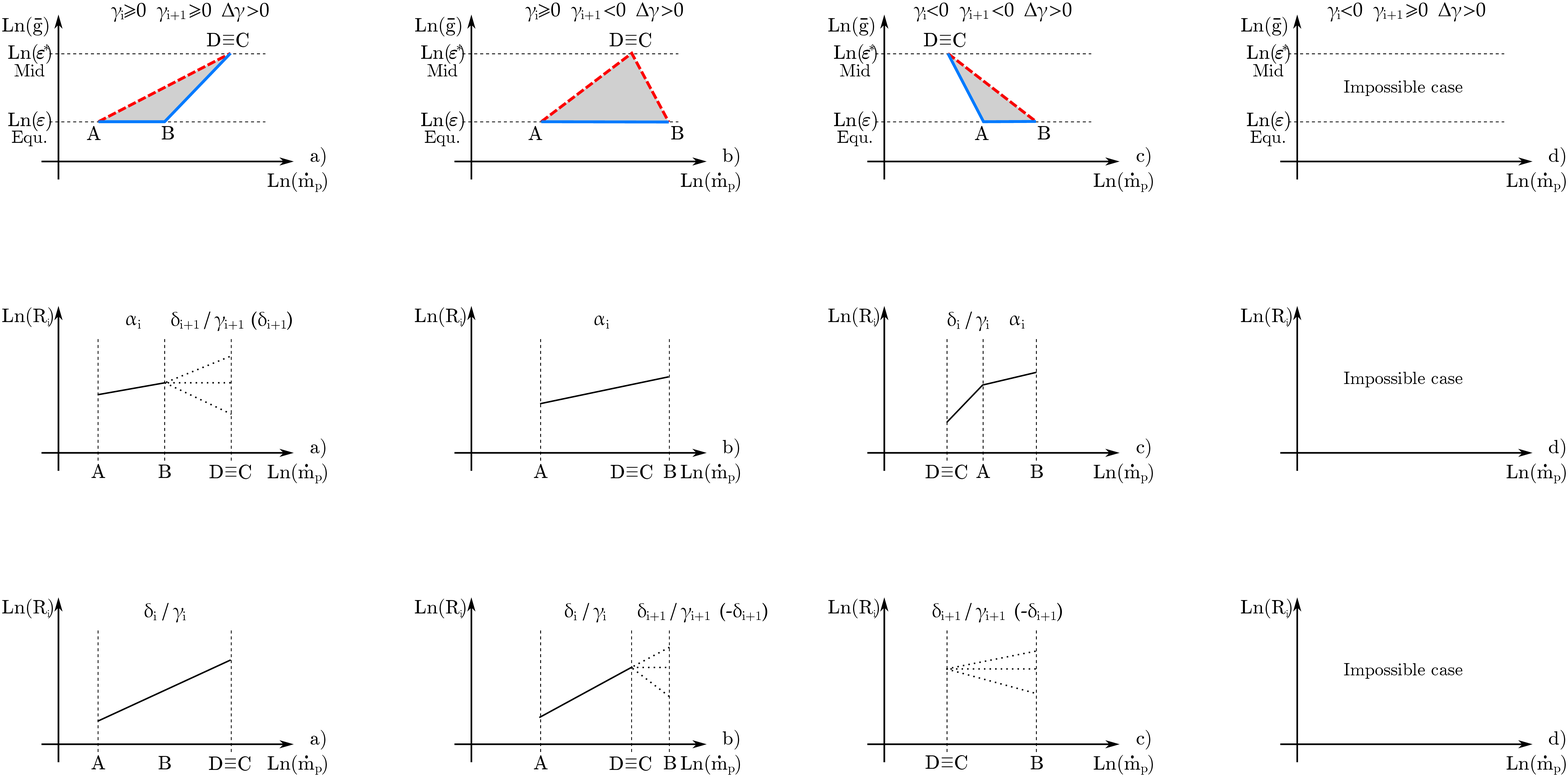}
    \caption{\uplp{Upper panel}: available colatitudes for different
      combinations of the sign of $\gami$ and $\gamip$ as in
      \figref{fig:TOS}, but for the case $\rhoi < 1$. One difference
      is that the maximum available for $\xbase$ is $\xsi$. Also, the
      fourth case, $\gami<0$ and $\gamip > 0$ (or $0$) is impossible
      due to the requirement $\delgami > 0$. The red dashed segments
      correspond to the \track{s} followed by ignition when $\beti >
      0$, the blue solid segments are the \track{s} followed when
      $\beti < 0$. When $\beti = 0$ any colatitude in the grey areas
      is equi-probable. \uplp{Middle panel}: burst rate evolution when
      $\beti < 0$, corresponding to the blue solid paths in the upper
      panel. As in \figref{fig:RBneg} we know that $\delii/\gami >
      \alpi > 0$ when it is the slope of the burst rate. The sign of
      $\deliip / \gamip$ is not known when needed and the slope sign
      is determined by the sign of $\deliip$. \uplp{Lower panel}:
      burst rate evolution when $\beti > 0$, corresponding to the red
      dashed paths in the upper panel. As in \figref{fig:RBpos} we
      know that $\delii/\gami > \alpi > 0$ when it is the slope of the
      burst rate. The sign of $\deliip / \gamip$ is not known and the
      slope sign is determined by the sign of $-\deliip$.}
    \label{fig:lessone}
  \end{figure*}
}

\newcommand{\biba}{
  \bibliographystyle{aasjournal}
  \bibliography{ms}
}

\newcommand{\cf}[1]{cf. \equlab{} \ref{#1}}

\newcommand{\thenu}{\left(\nu/\nuk\right)^2}
\newcommand{\bign}{N(\nu)}

\newcommand{\speed}{``speed''}
\newcommand{\norm}{$\xbase^\bet$}
\newcommand{\inc}{$\mdott^\alp$}

\newcommand{\KEP}{\texttt{KEPLER}~}

\shorttitle{Bursts rate, spin and accretion}
\shortauthors{Cavecchi et al.}

\begin{document}

\title{On the dependence of X-ray burst rate on accretion and spin
  rate}

\author[0000-0002-6447-3603]{Yuri Cavecchi}
\affiliation{Department of Astrophysical Sciences, Princeton University,
  Peyton Hall, Princeton, NJ 08544, USA}
\affiliation{Mathematical Sciences and STAG Research Centre,
  University of Southampton, SO17 1BJ, UK}

\author{Anna L. Watts}
\affiliation{Anton Pannekoek Institute for Astronomy, University of
  Amsterdam, Postbus 94249, 1090 GE Amsterdam, the Netherlands}

\author{Duncan K. Galloway}
\affiliation{School of Physics and Astronomy, Monash University, Clayton, VIC 3800, Australia}
\affiliation{Monash Centre for Astrophysics, Monash University, VIC 3800, Australia}

\correspondingauthor{Yuri Cavecchi}
\email{cavecchi@astro.princeton.edu}


\label{firstpage}

\begin{abstract}
  Nuclear burning and its dependence on the mass accretion rate are
  fundamental ingredients for describing the complicated observational
  phenomenology of neutron stars in binary systems. Motivated by high
  quality burst rate data emerging from large statistical studies, we
  report general calculations relating bursting rate to mass accretion
  rate and neutron star rotation frequency. In this first work we
  neglect general relativistic effects and accretion topology, though
  we discuss where their inclusion should play a role. The relations
  we derive are suitable for different burning regimes and provide a
  direct link between parameters predicted by theory and what is to be
  expected in observations. We illustrate this for analytical
  relations of different unstable burning regimes that operate on the
  surface of an accreting neutron star. We also use the observed
  behaviour of burst rate to suggest new constraints on burning
  parameters. We are able to provide an explanation for the long
  standing problem of the observed decrease of burst rate with
  increasing mass accretion that follows naturally from these
  calculations: when accretion rate crosses a certain threshold,
  ignition moves away from its initially preferential site and this
  can cause a net reduction of the burst rate due to the effects of
  local conditions that set local differences in both burst rate and
  stabilization criteria. We show under which conditions this can
  happen even if locally the burst rate keeps increasing with
  accretion.
\end{abstract}


\section{Introduction}
\label{sec:intro}

When a compact object with a solid surface such as a neutron star (NS)
is part of a binary system with a less evolved companion, accretion
onto the compact object may start which will lead to burning of the
fresh fuel accumulated on the surface of the NS. If the heating due to
the burning is not compensated by cooling, the burning will become
unstable, resulting in bright X--ray flashes: the so called type I
bursts \citep[see][]{rev-2003-2006-stro-bild-book}. A complete
description of the phenomenology of the observed bursts depends on
various factors, such as accretion physics
\citep{art-1999-ino-suny,art-2010-inoga-suny}, thermonuclear reaction
network physics
\citep{art-1981-fuji-han-miy,art-2000-cum-bild,art-2003-cum,
  art-2004-woos-etal,art-2007-heg-cumm-gal-woos,art-2016-cyb-etal} and
hydrodynamics that may regulate the flame propagation across the
surface following localised ignition \citep{art-2001-zing-etal,
  art-2011-malo-etal, art-2015-zing-etal, art-2013-cavecchi-etal,
  art-2015-cavecchi-etal,art-2016-cavecchi-etal}.

The implications of a complete understanding of the bursts go well
beyond the pure description of the thermonuclear flashes. Studying the
effects of the outcome of the burning can help in understanding the
structure of the compact object. For instance, in the case of NSs type
I bursts are one way to constrain the equation of state of the matter
in the inside \citep{rev-2013-miller-arxivpaper,rev-2016-watts-etal},
for example by inferring mass and radius from pulse profiles of burst
oscillations \citep[fluctuations in the lightcurves of the bursts due
  to asymmetric surface patterns that emerge during the bursts, see
  e.g.][]{rev-2012-watts}. Another example are the cooling lightcurves
of the NSs after the accretion outburst, which depend on how much (and
where) heat has been deposited by accretion and burning and also on
the structure of the outer layers of the star
\citep{art-1984-hana-fuji, art-1998-brown-bild-rut, art-2009-brow-cum,
  art-2013-wij-dege-page,art-2014-schatz-etal}, therefore providing a
very useful way of exploring NS (crust) properties such as
composition, structure, neutrino emission and superfluid
physics. Unfortunately, our understanding of the different ingredients
needed for modelling the observations is still limited. This paper
will discuss burning physics.

In the standard theoretical picture that emerges from calculations and
numerical simulations, how the burning proceeds depends on the burning
regimes (e.g. what fuel is available and what has been spent already,
which path the nuclear reactions follow, their temperature dependence
and their heat generation rate, see also \citeple{art-2011-schatz})
and the accretion rate. The accreted matter accumulates on the surface
of the star and sinks to deeper and deeper densities in the ocean,
eventually meeting the conditions where burning starts. At this point,
burning stability depends on whether the cooling is capable of
compensating the heat release or not. Even at low accretion rates the
burning rate and the energy release may be above the instability
threshold and the bursts begin; then the frequency of the bursts
increases with accretion rate. At the same time accretion releases
heat that eventually \emph{stabilises} the burning, \emph{preventing}
any bursting
\citep{art-1981-fuji-han-miy,art-1997-bild-brown,rev-1998-bild,art-2009-kee-lang-zand,
  art-2000-cum-bild,art-2014-zam-cumm-niq}. The amount of heat
generation from accretion comes from the gravitational energy released
at the moment of accretion, the compressional heat due to the extra
weight of the accumulated material and the heat of further reactions
that take place deeper than the burning layer
\citep{art-2000-cum-bild}. However, many details of the burst physics
are still uncertain, mainly the reaction rates
\citep[e.g.][]{art-2001-schatz-etal,art-2006-coop-nara-a,
  art-2006-coop-nara-b, art-2007-heg-cum-woos, art-2010-cyb-etal,
  art-2011-dav-cyb-etal, art-2014-keek-cyb-heger, art-2016-cyb-etal},
or, for example, the role of mixing \citep[e.g.][]{art-2007-piro-bild,
  art-2009-kee-lang-zand}.

One important factor in burst physics is the rotation of the
star. First of all, rotation opposes gravity, thus reducing the local
effective gravity, which has a direct effect on the \emph{local}
accretion rate and how the burning proceeds (for example determining
the most likely ignition colatitude,
\citealt{art-2007-coop-nara-a,art-2014-alge-mor} and see also next
\seclab{s}). Second, another source of heat that might have a
significant importance on the burning processes is the heat released
by some effective friction that takes place at the boundary and
spreading layers between the accretion disc and the surface of the
star \citep{art-1999-ino-suny, art-2010-inoga-suny,
  art-2014-kaja-etal, art-2016-philippov-rafi-stone}. For stars with
equal mass and radius, the magnitude of this effect will still depend
on the spin frequency of the star and how this compares to the
velocity of the disc at the star radius. Furthermore, rotation affects
the burning by inducing mixing of newly accreted material and ashes
from previous bursts in deeper layers. It also has indirect effects
since the mixing changes the temperature profile of the layer
\citep{art-2007-piro-bild, art-2009-kee-lang-zand}. Once again, the
exact dependence of bursting frequency on accretion rate and spin
frequency is still not well understood.

As a consequence of all the uncertainties, observations often do not
behave as models predict. For instance, the burning stabilises and the
bursts disappear too early, in terms of mass accretion rate, with
respect to theoretical expectations \citep[e.g.][but not always, see
  for example \citeple{art-2012-lin-etal}]{art-2003-corne-etal,
  art-2004-cum, art-2007-heg-cum-woos}. Also, most theoretical works
predict that the burst rate should always increase with accretion
rate. One notable exception is the delayed mixed burst regime found by
\citet{art-2003-nara-heyl}. However, \citet{art-2007-coop-nara-b}
caution against conclusions about the time dependent behaviour drawn
from linear stability analysis like the one of
\citet{art-2003-nara-heyl} and experimental work does not confirm the
prerequisite for the delayed mixed bursts \citep[namely a weaker CNO
  breakout reaction rate of $^{15}\rm{O}(\alpha,
  \gamma)^{19}\rm{Ne}$][]{art-2007-piro-bild, art-2007-fisk-etal,
  art-2007-tan-etal}. \citet{art-2007-coop-nara-a} also found a burst
rate decreasing with increasing accretion rate, but that was due to
the delayed burst regime of \citet{art-2003-nara-heyl}, which, as we
said, is not confirmed by direct
experiments. \citet{art-2016-lampe-heg-gal}, using the 1D multizone
code \KEP \citep{art-2004-woos-etal}, find a regime with decreasing
burst rate, but do so only in very limited ranges of high accretion
rate. Despite the fact that the general understanding would predict a
continuously increasing burst rate, the contrary is often observed: in
many sources the burst rate is seen to decrease by as much as an order
of magnitude before the bursts stabilize
(e.g. \citealt{art-1988-par-pen-lew, art-2003-corne-etal}, see also
\citealt{rev-2003-2006-stro-bild-book} and references therein); the
reason why is still not clear. Burst samples are now sufficiently
comprehensive \citep[e.g. the MINBAR catalogue,
  see][]{rev-2010-gal-etal} that we are able for the first time to
explore in a systematic way the effect of accretion and rotation rates
on the burst rate.

In this paper we present general, if somewhat simplified, calculations
which relate burning and accretion physics parametrizations to
observed quantities such as burst rate and mass accretion
rate\footnote{More precisely, mass accretion rate is not directly
  observed, but it is inferred from the X-ray luminosity under
  assumptions about the accretion flow and with some information on
  distance. However, as far as this paper and its calculations are
  concerned, we consider it in the category of `observables'.}. We
initially follow a similar approach to that of
\citet{art-2007-coop-nara-a}, who discussed the effects of neutron
star spin for a specific burning regime transitions. We develop the
calculations considering the effects of local gravity
(\secref{sec:definitions}) and we show how effects of mixing can be
included in the same formalism (\secref{sec:ignitionmp}). We present a
general study that covers all the mathematical possibilities and show
which ones would be compatible with observations. This paper is by
necessity leaning towards the abstract side, but we hope it would
offer a guide to the theoretical efforts and a bridge between theory
and observations.

\subsection{A new explanation for decreasing burst rate.}
\label{sec:summary}
  
The algebra of this paper will be presented fully in the following
sections, but since the mathematical steps may hide the physics and
the results behind them, we will discuss here the meaning and
implications of the calculations and how they compare with the
previous standing of the theory of bursts.

We will show that, by generalising the approach of
\citet{art-2007-coop-nara-a}, the burst rate of a single source can
be parametrized as (\cf{equ:buratebet}):

\begin{equation}
  \label{equ:sum:rate}
  \R = \Rb\, \mdot^{\alp}\, \xbase^{\bet}
\end{equation}
$\Rb$, $\alp$ and $\bet$ are constants that depend on the burning
regime. $\mdot$ is the local mass accretion rate at the pole, which
turns out to be a useful proxy for the global accretion rate
$\mdott$, as measured near the star, to which it is related
by (\cf{equ:mpmg}):

\begin{equation}
  \label{equ:sum:mpmg}
  \mdot = \frac{\left(\nu/\nuk\right) \sqrt{1 - \thenu}}
        {\arctan\sqrt{\frac{\thenu}{1 - \thenu}}}
        \frac{\mdott}{4\pi\Rs} = \bign \mdott
\end{equation}
where $\Rs$ is the radius of the star, $\nu$ the spin and $\nuk$ is
the Keplerian frequency at the star surface, $\nuk = \sqrt{G M /
  \Rs^3} / 2\pi$, so that

\begin{equation}
  \label{equ:sum:rate2}
  \R = \Rb \bign^{\alp}\, \mdott^{\alp}\, \xbase^{\bet}
\end{equation}
The important elements in \eqr{equ:sum:rate2} are \inc{} and \norm{.}
We have different forms for $\xbase$. $\xbase$ can be related to the
colatitude $\theta$ of the ignition by (\cf{equ:xbasedef})

\begin{equation}
  \label{equ:sum:xbase}
  \xbase = 1 - \left(\frac{\nu}{\nuk}\right)^2\sin^2\theta
\end{equation}
This equation expresses the correction to the local effective gravity
of the star at a given $\theta$ due to the centrifugal force. In
particular it expresses the ratio $\gefd{} / \gep$ of the local
gravity to the gravity at the pole. Later we will also suggest that
including the effects of mixing should give formulae of a similar form
to \eqr{equ:sum:rate2} (and \equlab{} \ref{equ:sum:stab} and
\ref{equ:therelation}) with $\xbase$ substituted by another function
of $\nu$ and $\theta$ that is also $1$ at the pole and $< 1$ at the
equator (see \secref{sec:mixing}). The effects of mixing should be
stronger than those due to changes in effective gravity, making mixing
a more plausible cause for the decreasing burst rate with accretion
rate; however, the argument for the mechanism we suggest could be
behind this phenomenon relies mostly only on the fact that there is a
dependence of burst rate on a function $\xbase$ which is greater at
the pole than at the equator.

The parameters $\alpha$ and $\beta$ in \eqr{equ:sum:rate2} depend on
the burning regime under consideration. $\alp$ clearly describes the
dependence on accretion rate, while $\bet$ is related to how the
ignition depth depends on the local effective gravity and on mixing,
which in turn are affected by the spin of the star and the colatitude
$\theta$, as noted above. In the most relevant cases, theory predicts
$\alp$ to be positive, which is also what intuition would predict: the
faster matter is accreted on the star, the faster the critical
conditions are reached for ignition. However, as already mentioned,
many sources show a complexity of different behaviours, most
importantly showing $\alp$ to apparently become negative (burst rate
decreasing) after some accretion rate $\mdott$.

Previously it was tentatively suggested that a possible reason for
that is a change in accretion geometry that leads to the local
accretion rate $\mdotgen$ to decrease while the global accretion rate
$\mdott$ increases \citep[see][and references therein]{rev-2000-bild,
  rev-2003-2006-stro-bild-book}, but exactly how this would happen was
not clear. \citet{art-2003-nara-heyl} and \citet{art-2007-coop-nara-a}
advocated instead a switch to their delayed mixed burst regime. We
provide a different explanation: the dimensionality of the problem is
the key.

Most of the burning physics theory is obtained with 1D simulations,
where the one dimension is the radial direction. While this approach
is extremely valuable, it does not take into account the fact that the
surface of the star adds two extra dimensions, namely $\theta$ and
$\phi$, where conditions are different even for a single star. The
role of $\xbase$ in \eqr{equ:sum:rate2} is then this: to incorporate
the effects of the second dimension $\theta$. $\xbase$ allows us to
take into account the fact that at \emph{different} colatitudes
$\theta$ of a \emph{spinning} NS the burst rate given by the
\emph{same} physics will be \emph{different} (basically due to the
different centrifugal force or different mixing). The importance of
this effect is given by the power $\bet$. It is difficult to find the
$\bet$ associated with the different theoretical works in the
literature, since this aspect is often neglected and the ignition
depth and its dependence on gravity and mixing are always reported
vaguely, if at all. However, we can extract it for example from the
calculations in \citet{rev-1998-bild} or
\citet{art-2007-piro-bild}. It can be seen that $\bet$ is expected to
be negative in the first case and positive in the second (see
\secrefs{sec:definitions} and \ref{sec:exabild} for more details).

\begin{figure}
  \centering
  \includegraphics[width=0.43\textwidth]{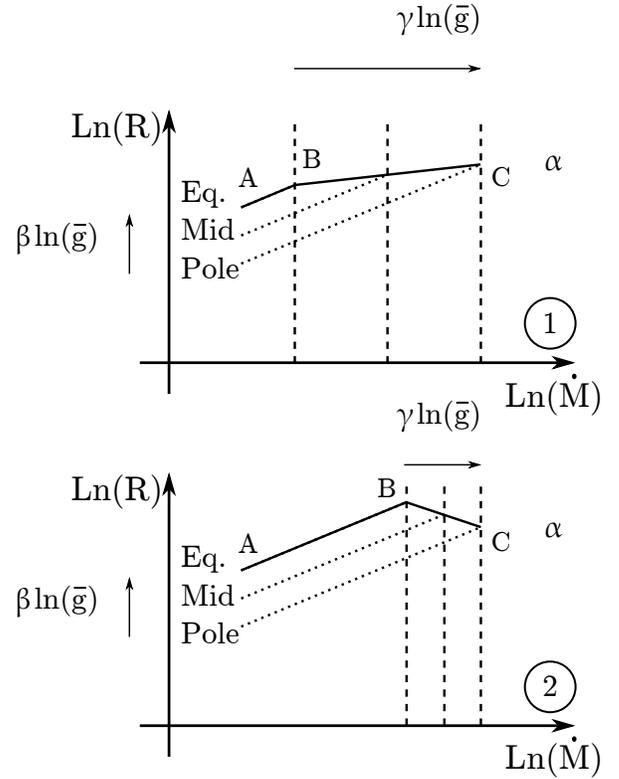}
  \caption{Burst rate $\R$ vs accretion rate $\mdott$, solid
    line. Example for $\alp>0$, $\bet<0$ and $\gamh>0$. The inclined
    dashed lines represent the burst rate at 3 colatitudes: equator,
    mid colatitude and pole. The slope is $\alp$. Equator has the
    advantage (bursts more often), since $\bet<0$. How much faster the
    burst rate of each colatitude is with respect to the pole is given
    by \norm{.} The three vertical dashed lines indicate the $\mdott$
    at which burning stabilizes at the various colatitudes. Ignition
    is highest at equator initially, but then it stabilizes and moves
    polewards. Depending on the \speed{} with which stabilization
    moves towards the pole, Case $1$ or $2$ are realized. The \speed{}
    of stabilization is given by $\Delta \ln \xbase / \Delta \ln
    \mdott = 1/\gamh$; $\gamh$ is the power with which the
    stabilization $\mdott$ depends on local conditions. Local
    conditions depend on colatitude $\theta$ and spin $\nu$. When the
    vertical lines are ``wide'', $\gamh$ is high, such that $\alp +
    \bet/\gamh >0$: the \speed{} is slow and the burst rate keeps
    growing, Case $1$. When the vertical lines are ``narrow'', $\gamh$
    is small such that $\alp + \bet/\gamh <0$: the \speed{} is high
    and the burst rate is seen to decrease, Case $2$.}
  \label{fig:sketch}
\end{figure}

So, how does this imply that above some critical $\mdott$ the burst
rate should decrease? The last ingredient to provide the answer is
the stabilization of burning. We show in \secref{sec:definitions}
that bursting is possible at any colatitude $\theta$ (this was also
recognised by \citealt{art-2007-coop-nara-a}) provided that

\begin{equation}
  \label{equ:sum:stab}
  \ml \xbase^{\gaml} \le \bign \mdott = \mdot \le \mh \xbase^{\gamh}
\end{equation}
where $\ml$ and $\mh$ are values dependent on the burning regime
(\cf{equ:mdotlimbet} and see \secref{sec:exabild} for an example).
$\ml \xbase^{\gaml}$ is the condition for the onset of
bursts\footnote{The limit for the onset of the bursts of a specific
  burning regime should be thought more accurately as the limit when
  the burst rate of that specific regime becomes faster than the rate
  of the other regimes.}, $\mh \xbase^{\gamh}$ is the condition for
stabilization. As for $\bet$, $\gaml$ and $\gamh$ are related to the
ignition depth and its dependence on the local effective gravity or
mixing. Once again it is difficult to obtain values of $\gamh$ from
the literature, but again we can infer its value for the cases treated
by \citet{rev-1998-bild}, where $\gamh > 0 $, or
\citet{art-2007-piro-bild}, where $\gamh < 0$. Note however that there
is uncertainty around these values (see \secref{sec:exabild}).

The explanation we suggest for the decreasing burst rate then goes as
follows (see \figref{fig:sketch} for a sketch). Let's consider the
case $\bet < 0$, $\gamh > 0$. Initially, the most probable ignition
location is the equator, point \Ap{} on \figref{fig:sketch}, because
$\bet<0$, $\xbase(\theta=\pi/2) < \xbase(\theta=0)$ and this makes the
rate at the equator the highest, \eqr{equ:sum:rate2}.  With increasing
$\mdott$ the most probable ignition site will remain on the equator,
until condition \eqr{equ:sum:stab} is broken (point \Bp{}). In the
range \segt{} of accretion rate the burst rate should be increasing as
$\mdott^{\alp}$ because the factor $\xbase^{\bet}$ in
\eqr{equ:sum:rate2} will not change. The fact that ignition stays on
the equator depends on the fact that $\bet < 0$ (see
\secref{sec:ignitionmp} for further details and more
possibilities). After point \Bp{,} while the accretion rate $\mdott$
increases, the most probable ignition colatitude moves towards the
pole, while the part near the equator should be burning stably. When
the pole becomes the most probable location, point \Cp{,} the whole
star should be burning stably and the bursts should disappear. In the
range \segd{} the rate of the bursts will go as

\begin{equation}
  \label{equ:therelation}
  \Rb \propto \mdott^{\alp + \bet/\gamh}
\end{equation}
Depending on the sign of $\alp + \bet/\gamh$ the burst rate may
actually decrease.

Note that \emph{this condition is NOT in contradiction with the
  theoretical results of simulations that give consistently increasing
  bursting rate as a function of $\mdott$}. As can be seen in
\figref{fig:sketch}, at a fixed colatitude $\xbase$ is a constant and
the rate \emph{is} increasing as a function of $\mdott$, but the
dependence of ignition depth and burst rate on \emph{local position},
measured by $\xbase^{\bet}$, makes the normalization factor in
\eqr{equ:sum:rate2} different at different colatitudes. The
normalization is higher at the equator (if $\bet<0$), so that the
overall burst rate (normalization) near the pole can be significantly
lower than at the equator.

If the \speed{} in terms of $\mdott$ at which ignition moves polewards
is fast enough, the increase in burst rate due to \inc{} will not be
able to compensate for the initial deficit due to the normalization
factor \norm{} and the burst rate will decrease. The \speed{} at which
the ignition moves polewards can be thought of as $\Delta \theta /
\Delta \mdott$ or, more conveniently, $\Delta \ln \xbase / \Delta \ln
\mdott = 1/\gamh$, see \eqrs{equ:sum:stab} and \eqref{equ:gii}. Small
$\gamh$ leads to high \speed{} and decreasing burst rate, Case $2$ in
\figref{fig:sketch}. High $\gamh$ gives slow \speed{} and the increase
\inc{} is able to cover the gap due to normalization factor \norm{}
and the observed burst rate will increase. We discuss more the role of
$\alp$, $\bet$ and $\gam$ in \secref{sec:discussion}.

Finally, note that the fact that ignition moves off its initial site
due to stabilization may also explain why bursts at high accretion
rate seem to be less energetic \citep{art-1988-par-pen-lew}. A smaller
fraction of the star surface would be burning efficiently, since part
of the fuel in the stabilized regions will have been spent in stable
burning\footnote{Of course this is similar to the suggestion of
  \citet{art-2003-nara-heyl}, but here the origin of the stable
  burning is not the delayed mixed burst regime, it is the competition
  between $\bet$ and $\gamh$ in the power of \eqr{equ:therelation}. In
  this sense, the explanation is more similar to \citet{rev-2000-bild}
  even though we do not invoke any strongly changing accretion
  geometry.}. In the case of equatorial ignition, this very same
mechanism may help explain why bursts seem to stabilize before the
expected $\mdott$: the theoretical $\mdott$ from 1D multizone
simulations is the one corresponding to conditions at the pole, point
\Cp{} since corrections due to rotations are absent there. However, at
that point the bursts may have become too weak and rare to be
detected.

\section{The relation between bursting rate, accretion and spin frequency}
\label{sec:definitions}

We begin by generalizing and extending the approach of
\citet{art-2007-coop-nara-a}. Thus, we initially present results
regarding the local effective gravity. In \secref{sec:mixing} we argue
that mixing can have effects on the burst rate that are
\emph{formally} very similar to the effective gravity, even though of
different magnitude. Mixing has not been explored as thoroughly as
gravity has. The latter offers therefore a more solid ground for
beginning this presentation. The burning rate of a specific regime is
generally described as a function of effective gravity $\gfed$ and
local accretion rate $\locmdot$ \citep[where $\nu$ is the spin
  frequency and $\theta$ is the colatitude measured from the north
  pole, see for example][]{rev-1998-bild,rev-2000-bild}. Without
considering general relativistic corrections\footnote{For the effects
  of general relativity see \citet{art-2014-alge-mor} and
  \secref{sec:discussion}.}, $\gfe$ is written

\begin{equation}
  \label{equ:geff}
  \gfe = g - \Omega^2 \Rs \sin\theta^2
\end{equation}
where $\Omega$ is the angular velocity of the star ($\Omega = 2 \pi
\nu$) and $\Rs$ is the radius of the star. If we take $\gep = g = G M /
\Rs^2$, we can write

\begin{equation}
  \label{equ:xbaseeff}
  \gfe = \gep \left[1 - \left(\frac{\nu}{\nuk}\right)^2\sin^2\theta\right]
\end{equation}
where we have introduced the Keplerian frequency $\nuk = \sqrt{G M /
  \Rs^3} / 2\pi$, $G$ is the gravitational constant and $M$ is the
mass of the star. We will write $\gfe = \gep\, \xbased$ for later
convenience, so that
\begin{equation}
  \label{equ:xbasedef}
  \xbase = 1 - \left(\frac{\nu}{\nuk}\right)^2\sin^2\theta
\end{equation}
$\xbase$ depends on spin and colatitude, but also on the mass and
radius of the star through $\nuk$. It is the ratio $\gfed{} / \gep$
and it measures the \emph{modification} to local gravity due to
rotation with respect to a non rotating star. \emph{Presently it has
  to be interpreted as a function of position $\theta$ (and $\nu$)}.

We also introduce the number $\ep$, which is $\xbase$ evaluated at the
equator,
\begin{equation}
  \label{equ:ep}
  \ep = \xbase(\pi/2, \nu) = 1 - (\nu/\nuk)^2
\end{equation}
so that $\ep \le \xbased \le 1$. $\ep$ is a quantity characteristic of
each specific neutron star, combining spin frequency, mass and radius
of the star.  It is equal to $1$ for non rotating stars and equal to
$0$ for stars rotating at the Keplerian frequency. This latter limit
is non physical because the star would not be bound, at least at the
equator.

Assuming that the accreted material spreads rapidly over the surface,
the local accretion rate $\locmdot$ at a specific colatitude is
related to the local accretion rate at the pole $\mdot$ by
\citep{art-2007-coop-nara-a}

\begin{equation}
  \locmdot = \mdot \gep \gefd{-1} = \mdot \xd{-1}
  \label{equ:mdrel}
\end{equation}
We note that the local accretion rate at the pole can be related to
the global accretion rate $\mdott$, the total amount of mass accreted
per unit time as measured near the star, or to the surface-averaged
local accretion rate $\mdotg$ as follows.

\begin{figure}
  \centering
  \includegraphics[width=0.45\textwidth]{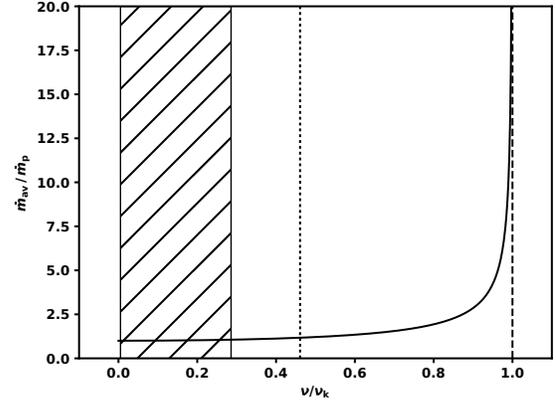}
  \caption{The relation between the \emph{observed, average} accretion
    rate $\mdotg$ and the \emph{local} accretion rate \emph{at the
      pole} $\mdot$. See \eqr{equ:mpmg} and note that $\nu/\nuk =
    \sqrt{1 - \ep}$. The divergence of the ratio $\mdotg/\mdot$ when
    $\nu = \nuk$, vertical asymptote (shown by the dashed line), is
    due to the fact that for a star rotating at the Keplerian
    frequency the local accretion should be $0$. This plot is general
    and as a specific example the dotted line indicates the position
    of a star of $M = 1.4$ \ms and $\Rs = 10$ km spinning at $10^3$
    Hz. The hatched region indicates the range of the known bursters:
    $11$ Hz \citep{atel-2010-alta-etal} - $619$ Hz
    \citep{art-2003-hart-etal}.}
  \label{fig:mav}
\end{figure}

$4 \pi \Rs^2 \mdotg = \mdott = \int \locmdot \Rs^2 \sin\theta \rd
\theta \rd \phi$, where the integral is extended over the whole
surface (assumed to be of spherical shape for simplicity and
consistency with \equlab{.}  \ref{equ:geff})\footnote{If we were to
  include the effects of oblateness, the integral would be $\mdott =
  \int \locmdot \Rs^2 \left[1 + f^2(\theta)\right]^{1/2} \sin\theta
  \rd \theta \rd \phi$, where $f(\theta) = (\rd \Rs / \rd \theta) /
  \Rs$ and $\Rs$ is a function of $\theta$ only. This effect should be
  relevant only for very fast rotating stars \citep[$\nu/\nuk \gtrsim
    0.3$][]{art-2014-alge-mor}, unless general relativity effects are
  taken into account.}. With \eqrs{equ:xbasedef} and \eqref{equ:mdrel}
this leads to $\mdotg = \mdot / 2 \int \xd{-1} \sin\theta \rd \theta$,
where the integral in $\phi$ yields $2\pi$ since $\xbase$ does not
depend on $\phi$. After some algebra the result is

\begin{equation}
  \label{equ:mpmg}
  \frac{\mdott}{4 \pi \Rs^2} =
  \mdotg = \mdot \frac{\arctan\sqrt{(1-\ep)/\ep}}
         {\sqrt{\ep (1 - \ep)}} 
\end{equation}
$\ep=0$ is not admissible and for $\ep \to 1$ $\mdot\to\mdotg$. The
mapping between $\locmdot$, $\mdot$ and $\mdotg$ should be taken into
account when comparing to observations, since observations are usually
stated in terms of $\mdott$ or $\mdotg$, while theoretical models
prefer the use of $\locmdot$. \eqrs{equ:mdrel} and \eqref{equ:mpmg}
show how the relation between these quantities depends on the rotation
frequency and the mass and radius of each star and is therefore
different for different systems (see also \figref{fig:mav}). However,
for a star of mass $\Ms = 1.4$ \ms and $\Rs = 10$ km rotating at $\nu
= 10^3$ Hz ($\nu/\nuk \approx 0.43$, $\ep \approx 0.82$) the
correction due to \eqr{equ:mpmg} is only $1.14$, so that this
correction becomes important only for very rapidly rotating systems.

Analytical calculations show that the ignition depth $y_{\rm{ign}}$
(the column density in g cm$^{-2}$ at which ignition takes place) can
be expressed as a function of local mass accretion rate, local gravity
and the properties of the burning regime under consideration, and this
is confirmed by direct numerical experiments \citep[see][and also
  \secref{sec:exabild}]{art-1981-fuji-han-miy,rev-1998-bild}. In
general, expressions for the ignition temperature and depth can be
estimated combining the equations for the temperature profile across
one column of fluid, obtained for example under the assumption of
constant flux, to the conditions for unstable burning and/or depletion
of a specific species \citep{art-1981-fuji-han-miy,rev-1998-bild}.
The flux depends on the burning regime or on extra heat sources,
usually proportional to the accretion rate, like gravitational energy
release or extra nuclear reactions at the bottom of the ocean. The
conditions for instability are obtained comparing energy release rate
due to burning and cooling rate. Gravity enters the equations also
through the equation of state of the burning fluid and the relation
between pressure $P$ and column depth: $P = y \gfe$. Then, the burst
recurrence time can be expressed as the time it takes for accreted
fluid to reach the ignition depth: $t_{\rm{rec}} = \yig /
\locmdot$ \citep{rev-1998-bild,art-2007-coop-nara-a} and therefore we
can write (see \secref{sec:exabild} for an explicit example)

\begin{equation}
  \label{equ:trec}
  t_{\rm{rec}} \propto \locmdot^{-\alpt}\,\gefd{-\bett}
\end{equation}
This expression could be used also to fit measurements from numerical
experiments, therefore making it even more generally useful.

The bursting rate $\R$ is the inverse of the recurrence time, which
leads to

\begin{equation}
  \label{equ:protorate}
  \R \propto \locmdot^{\alpt}\,\gefd{\bett}
\end{equation}
This can be re written as

\begin{equation}
  \label{equ:buratebad}
  \R = \Rb\, \locmdot^{\alpt}\,\xd{\bett}
\end{equation}
$\Rb$ is a \emph{pseudo} constant that includes dependence on the mass
and radius through $\gep = G M / \Rs^2$ and physical parameters like the
fluid composition and conductivity \citep[see the example of
  \secref{sec:exabild}, where we apply this to \alienequlab{s.} 20 and
  32 in][and remember that $\R = \locmdot /
  y_{\rm{ign}}$]{rev-1998-bild}. Using \eqr{equ:mdrel}, we can write

\begin{equation}
  \label{equ:buratebet}
  \R = \Rb\, \mdot^{\alp}\,\x{\bet}
\end{equation}
where $\alp = \alpt$, $\bet = \bett - \alpt$. In order to avoid
cumbersome notation we have dropped the explicit dependence over
$\theta$ and $\nu$ from $\xbase$, but that should be kept in mind
since the role of $\xbase$ is to track the colatitude.

Typically, the bursting rate of a specific burning regime is only
valid within an interval of local mass accretion rate, outside of
which either the burning is stable or the burst rate of another regime
is higher. The limits for stability set conditions on the burning
temperature which, being found in a similar way to $y_{\rm{ign}}$, can
be expressed in terms of $\mdotgen$ and $\gfe$ \citep[see for example
  the derivation of \alienequlab{s.} 24 - 26 or 36
  of][]{rev-1998-bild}. The precedence of one regime over another is
mainly set by comparing the column depth $y_{\rm{ign}}$ at which
different regimes ignite and checking which one is smaller; once again
these conditions involve $\mdotgen$ and $\gfe$
\citep[e.g. \alienequlab{.} 35 of][]{rev-1998-bild,
  art-2007-coop-nara-a}.  As a consequence, these limits are quite
generally of the form

\begin{equation}
  \label{equ:mdotlimbad}
  \ml \x{\gamtl} \le \locmdot \le  \mh \x{\gamth} 
\end{equation}
Where $\ml$, $\mh$ are again \emph{pseudo} constants that hide
dependence on physical parameters in the same way as $\Rb$. The
$\gamt$s are parameters that depend on the burning regimes and
$\gamtl$ need not necessarily be equal to $\gamth$ (see an example in
\secref{sec:exabild}). As for the burst recurrence time, these
expressions could be used to fit the results from numerical
simulations, thus providing a useful general form.

Thanks to \eqr{equ:mdrel} these constraints can again be written for
convenience as
\begin{equation}
  \label{equ:mdotlimbet}
  \ml \x{\gaml} \le \mdot \le  \mh \x{\gamh} 
\end{equation}
where $\gam_* = \gamt_* + 1$. Forms \eqref{equ:buratebet} and
\eqref{equ:mdotlimbet} are preferable over \eqref{equ:buratebad} and
\eqref{equ:mdotlimbad} respectively because they express the two
conditions in such a way that the dependence over $\theta$ and $\nu$
(or $\ep$) is only present through $\xbase$ and clearly separated from
the dependence on the accretion rate, which is parametrized by
$\mdot$. $\mdot$ has to be interpreted as a parameter that acts as a
proxy for the \emph{observational information} $\mdotg$ ($\mdott$):
the link being provided by \eqr{equ:mpmg}. In principle
\eqrs{equ:buratebet} and \eqref{equ:mdotlimbet} could be expressed
directly in terms of $\mdotg$ and $\nu$ (or $\ep$), but that would
make the following equations even more cumbersome.

Finally, a star can experience different dominant burning regimes, so
we shall write in general

\begin{equation}
  \Ri = \Rbi\, \mdot^{\alpi}\,\x{\beti}
\end{equation}
and
\begin{equation}
  \label{equ:mdotrange}
  \mi \x{\gami} \le \mdot \le  \mip \x{\gamip} 
\end{equation}
where the index $i$ indicates the burning regime. The critical
accretion rate $\mip\x{\gamip}$ is also the lower limit of rate $\Rip$
etc.

The last quantities we need to define are the burst ignition rate
evaluated at the two $\mdot$ extremes of applicability:

\begin{IEEEeqnarray}{lll}
  \label{equ:rii}
  \Rii  &=\Ri|_{\mdot=\mi \x{\gami}}  &=\, \Rbi\, \mi^{\alpi} \,\x{\delii}\\
  \label{equ:riip}
  \Riip &=\Ri|_{\mdot=\mip \x{\gamip}}&=\, \Rbi\, \mip^{\alpi}\,\x{\deliip}
\end{IEEEeqnarray}
where
\begin{align}
  \label{equ:delii}
  \delii =& \,\alpi \gami + \beti \\
  \intertext{and}
  \label{equ:deliip}
  \deliip =& \,\alpi \gamip + \beti.
\end{align}
are useful shortening notation (note also that $\del_{\ii,*} =
\alpt_{\ii}\gamt_{*} + \bett_{\ii}$).

\section{Where does ignition take place, given a specific $\mdot$?}
\label{sec:ignitionmp}

For a given star with given gravity and spin frequency, ignition is to
be expected at the colatitude where the rate is higher
\citep{art-2007-coop-nara-a}. Let's consider the regime $i$. The first
question is whether \emph{at each colatitude} the regime can be
realized at all. From \eqr{equ:mdotrange} we can see that at each
$\theta$ we need $\mi \x{\gami} \le \mip \x{\gamip}$. Otherwise the
regime $i$ would be skipped there in favour of regime $i + 1$ (or $i -
1$). This translates into

\begin{equation}
  \label{equ:burnexists}
  \x{(\gami-\gamip)} \le \left(\frac{\mip}{\mi}\right)
\end{equation}
It will be useful to define
\begin{equation}
  \label{equ:delgami}
  \delgami = \gami - \gamip \;(= \gamt_{\ii} - \gamt_{\iip})
\end{equation}
\begin{equation}
  \rhoi = \frac{\mip}{\mi} (\ge 1)
\end{equation}
and
\begin{equation}
  \label{equ:epsstar}
  \epsstari = \rhoi^{1/\delgami}
\end{equation}
It is easy to see that \eqr{equ:burnexists} is satisfied by
\begin{IEEEeqnarray}{r?t?l}
  \label{equ:exist1}
  \ep  \le \xbase \le 1 & if &
                          \delgami < 0 \anda \epsstari \lt \ep
                          \trextra{or} \delgami \ge 0.\\
  \label{equ:exist2}                        
  \xsi \le \xbase \le 1 & if &
                          \delgami < 0 \anda \epsstari \ge \ep.
\end{IEEEeqnarray}
Note that \eqrs{equ:exist1} and \eqref{equ:exist2} show that
$\epsstari$ marks a critical value for $\ep$, and therefore for $\nu$,
across which the behaviour switches in the case of $\delgami < 0$. If
$\rhoi$ would be allowed to be also $\rhoi < 1$, there would exist
cases where the maximum possible $\xbase$ would be less then one, i.e.
ignition may not reach the pole, in analogy to the cases where the
minimum value is $\epsstari$ and not $\ep$ (i.e. some mid latitude and
not the equator). That $\rhoi < 1$ seems highly unlikely, and
therefore we do not treat this extra possibility here; see however
\appref{sec:lessone}. The colatitude $\thesi$ which corresponds to
$\epsstari$ is given by

\begin{equation}
  \label{equ:thesi}
  \thesi = \arcsin\sqrt{\frac{1 - \rhoi^{1/\delgami}}{1 - \ep}}
\end{equation}
$\thesi$ is the solution of $1 - (\nu/\nuk)^2 \sin^2\thesi = \xsireal
= \epsstari$ and corresponds to $\pi/2 - \lambda_{\rm{ign}}$ of
\alienequ{8b} of \citet{art-2007-coop-nara-a}. There exists also the
solution $\pi - \thesi$, but this is in the southern hemisphere. Since
the north and south hemispheres are symmetrical, we consider only
north hemisphere solutions. The condition for the existence of
$\thesi$, if $\rhoi \ge 1 $ is the same as \eqr{equ:exist2}.

\eqrs{equ:exist1} and \eqref{equ:exist2} establish the range of
colatitudes (parametrized by $\xbase$) where bursts can happen. The
next question is: at a given accretion rate, parametrized by $\mdot$,
where does ignition take place first, among the allowed colatitudes?
This question was addressed by \citet{art-2007-coop-nara-a} and we
present its generalisation here.

\subsection{Another mechanism affecting the burst rate: mixing}
\label{sec:mixing}

This is a good place to introduce another physical mechanism that
affects burst rate, regime switching and stability. In our formalism
that means another form for $\xbase$. In the derivation so far we
followed \citet{art-2007-coop-nara-a} and used the effects of the
centrifugal force on local gravity to identify a function $\xbase$
that would have the following properties: 1) depends on spin and
latitude (being $1$ at the pole and $<1$ at the equator) and 2)
changes the local behaviour of bursts. The centrifugal force case is
more intuitive being well know from the literature. However, another
mechanism that depends on spin and is known for affecting the burst
behaviour is mixing. \citet{art-2007-piro-bild} give analytical and
linear stability analysis results about mixing, in particular mixing
due to the effective viscosity resulting from the Tayler-Spruit dynamo
\citep{art-1999-spruit, art-2002-spruit}. The authors found that the
mixing was more effective for slowly rotating
stars. \citet{art-2009-kee-lang-zand} performed more sophisticated,
yet still 1D numerical simulations showing that mixing could be
important also for fast spins. They also found that mixing due to
other, purely hydrodynamical effects could be important for high
enough spins.  However, they did not provide analytical expressions.

The analytical formulae of \citet{art-2007-piro-bild} are particularly
useful for this paper, since they express the burst rate as $\R
\propto \mdotgen{}^{\alp} \nu^{-\bet}$ and the limits for burning
regimes as $\mdotgen_{\rm{crit}} \propto \nu^{-\gam}$. These formulae
are derived based on equations averaged over the surface, especially
over $\theta$, but some dependence over $\theta$ is to be expected in
reality \citep[see][]{art-1993-fuji, art-2002-spruit}. Finding the
exact formulae is beyond the scope of this paper, even though it
definitely warrants further work based on the conclusions of
\secref{sec:discussion} (see also \secref{sec:summary}) where we
suggest that they could provide an explanation for the decreasing
burst rate.

We can \emph{speculate}, however, just in order to give a concrete
example of what we mean. The biggest difficulty is how to extend the
formulae of \citet{art-2002-spruit} and \citet{art-2007-piro-bild} to
the entire surface of the star keeping the dependence over $\theta$
explicit. Since the Tayler-Spruit dynamo depends on an external source
to keep the shear in the vertical direction and this can be provided
more easily near the equator by the accretion disc, the simplest
possibility is to consider something like $\nu\sin\theta$. Note also
that this formulation becomes unphysical, predicting infinite (or at
least very high) rates for slow rotators. We can speculate on the
existence of a limiting, perhaps very small, value $\num$ such that we
can write for example $\R \propto (\num + \nu\sin\theta)^{-\bet}
\propto(1 + \nu\sin\theta/\num)^{-\bet}$. The exact form is not
important here, but it should be investigated for seeking more
quantitative analysis of course: we use this one only as an
\emph{example}. Then one can write

\begin{equation}
  \label{equ:xbasem}
  \xbasem = \left(1 + \frac{\nu\sin\theta}{\num}\right)^{-1}
\end{equation}
The reason for the negative power is that in this way $\xbasem$ will
be one at the pole and take a value $\epm$ at the equator, just like
$\xbase$ of \eqr{equ:xbasedef}. In this \emph{example} the value at
the equator would be

\begin{equation}
  \label{equ:epm}
  \epm = \left(1 + \frac{\nu}{\num}\right)^{-1} < 1
\end{equation}
again characteristic of each star. Finally, for a given $\epsstari$,
the corresponding colatitude would be

\begin{equation}
  \label{equ:themsi}
  \themsi = \arcsin \frac{{\epsstari}^{-1} - 1}{{\epm}^{-1} - 1}
\end{equation}

We do not discuss this formulation anymore, because it is not the goal
of this paper, but it is not unreasonable to think that a similar
expression to \eqr{equ:xbasem} actually takes place. From such a
formula, definitions for $\epm$ and $\themsi$ could be obtained as we
did for our example.

As a final remark we note that if the effects of mixing are taken into
account, the full formulae should in principle still include the
effects of gravity: $\R \propto \mdot^{\alp}\, \x{\bet}\,
\xm{\betm}$. However, since $\gefbase$ does not change much from pole
to equator, the effects of mixing should be dominant, unless the
dependence over gravity is much higher than presently understood. The
change over gravity could be thus neglected. From now on, we will only
write our discussion in terms of $\xbase$, $\ep$ and $\thesi$. The
same conclusions apply to the functions set by gravity and the
centrifugal force as in \eqrs{equ:xbasedef}, \eqref{equ:ep} and
\eqref{equ:thesi} or to the functions set by mixing, as in our example
\eqrs{equ:xbasem}, \eqref{equ:epm} and \eqref{equ:themsi}.

\subsection{Ignition latitude of type I bursts}
\label{sec:ignlat}

\begin{figure*}
  \centering
  \noindent
  \twofiginarow{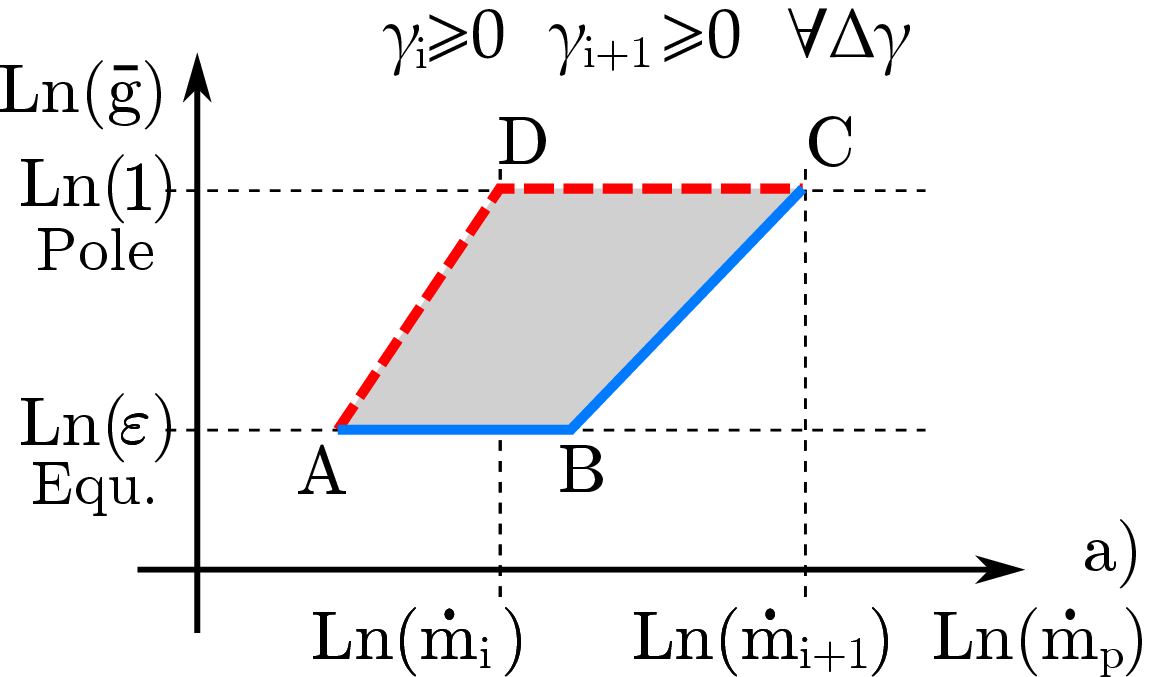}{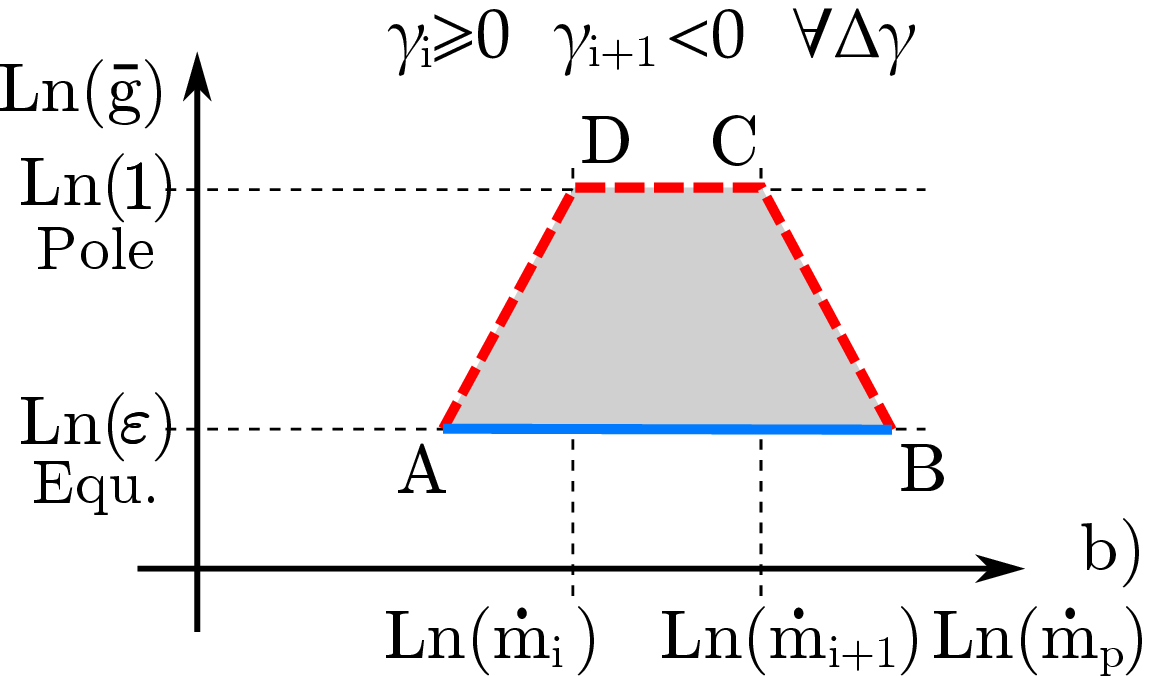}{0.3}\\\vspace{\baselineskip}
  \twofiginarow{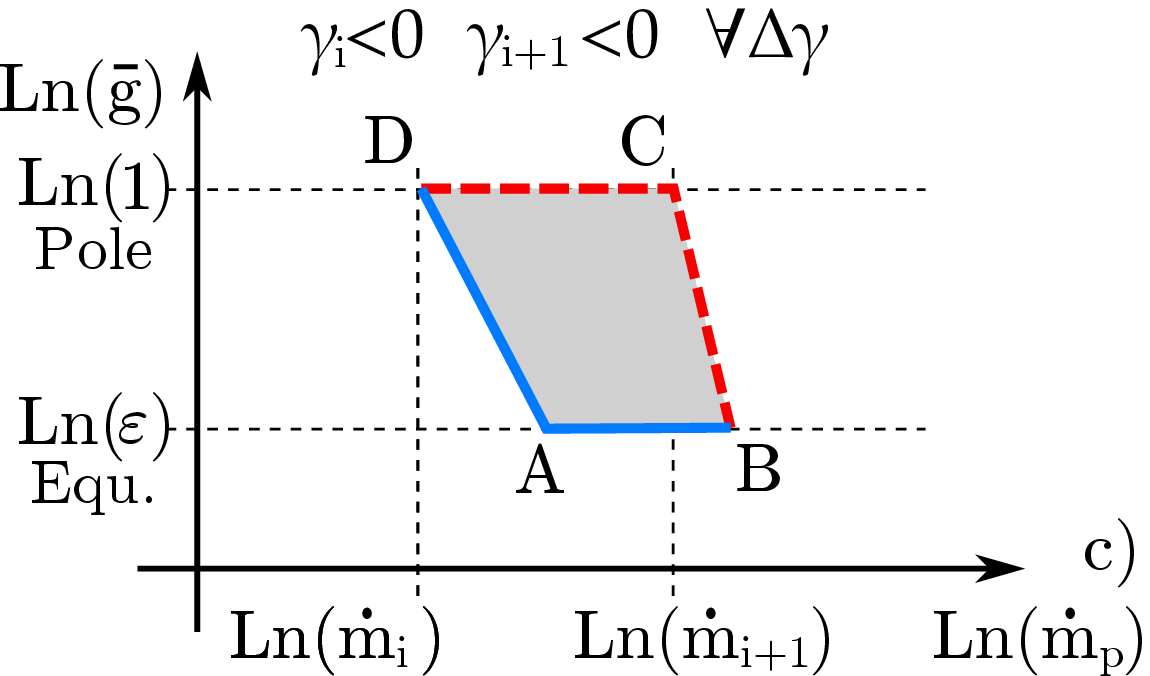}{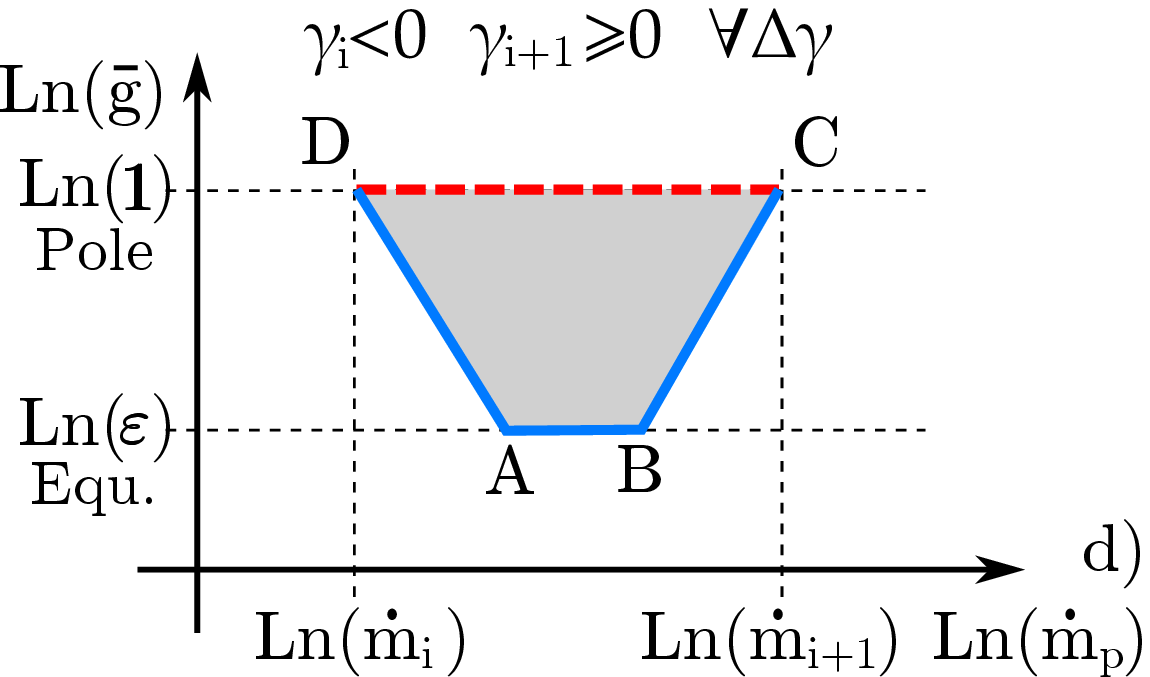}{0.3}\\\vspace{\baselineskip}
  \twofiginarow{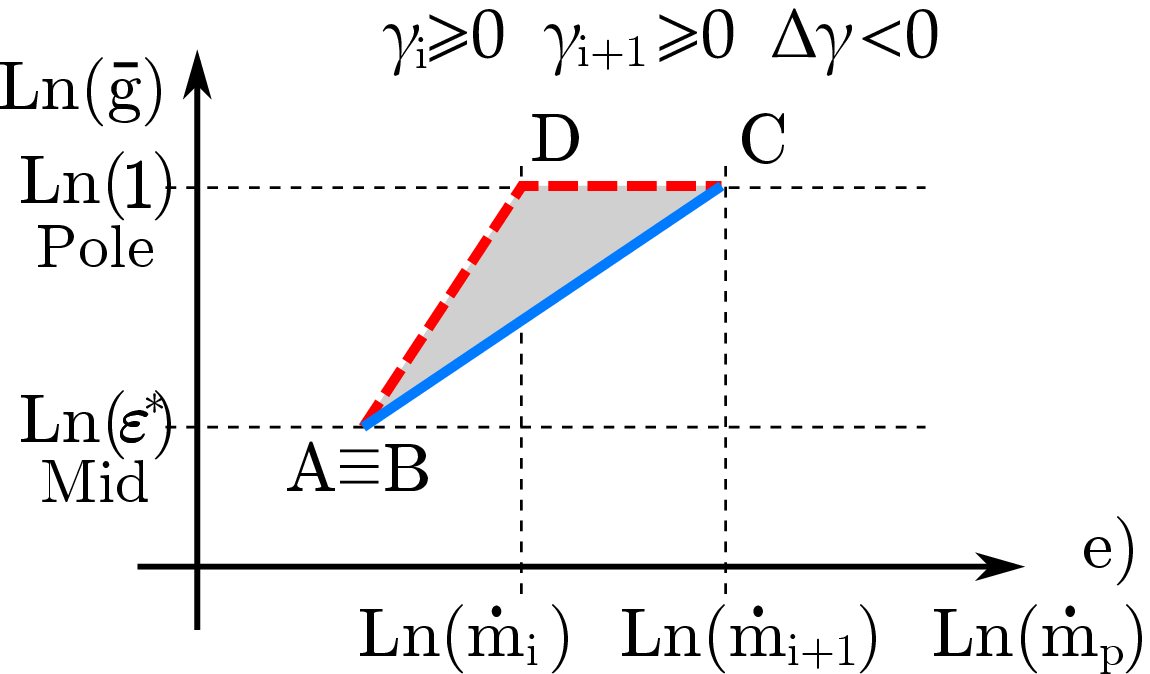}{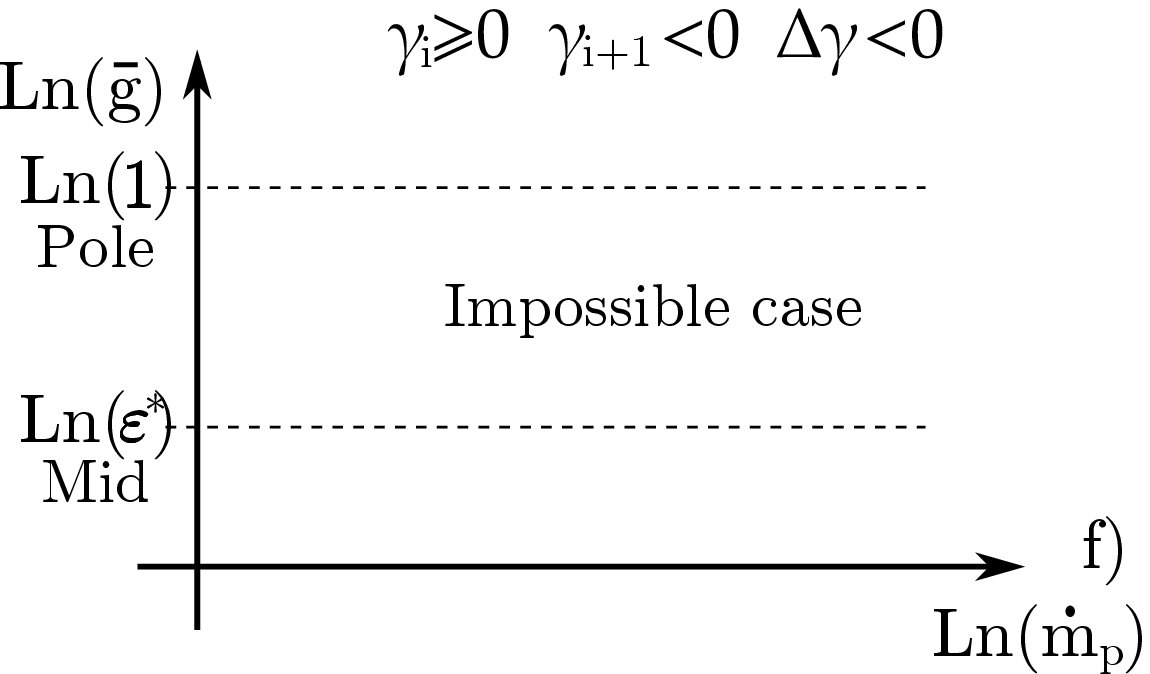}{0.3}\\\vspace{\baselineskip}
  \twofiginarow{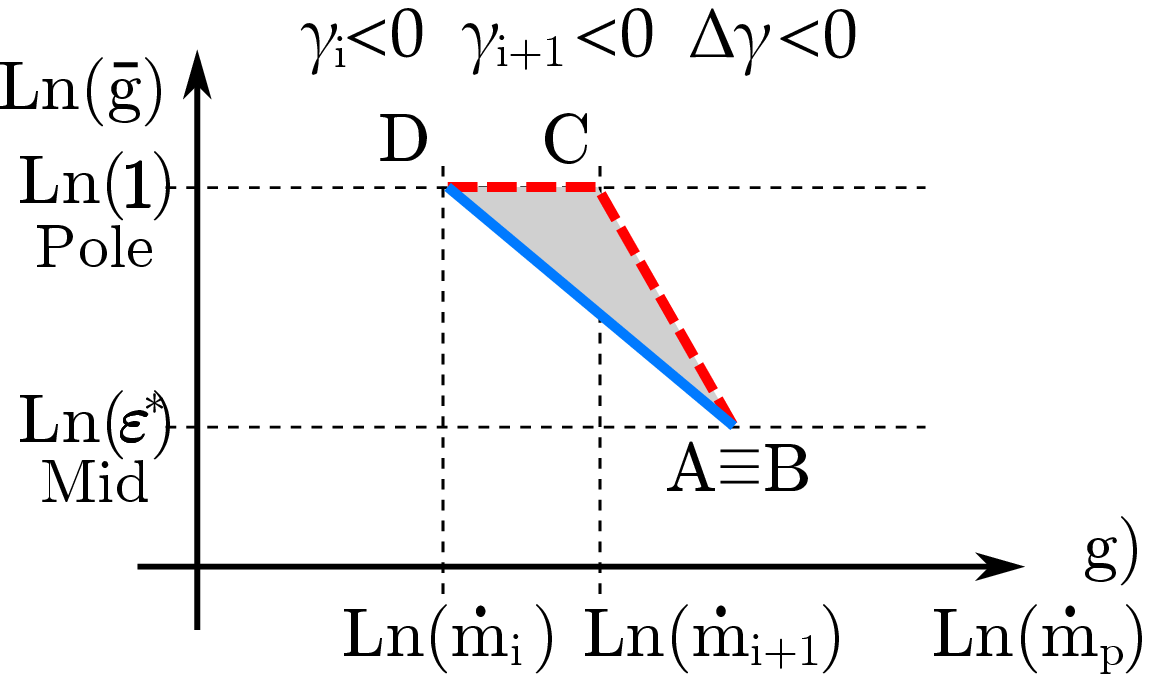}{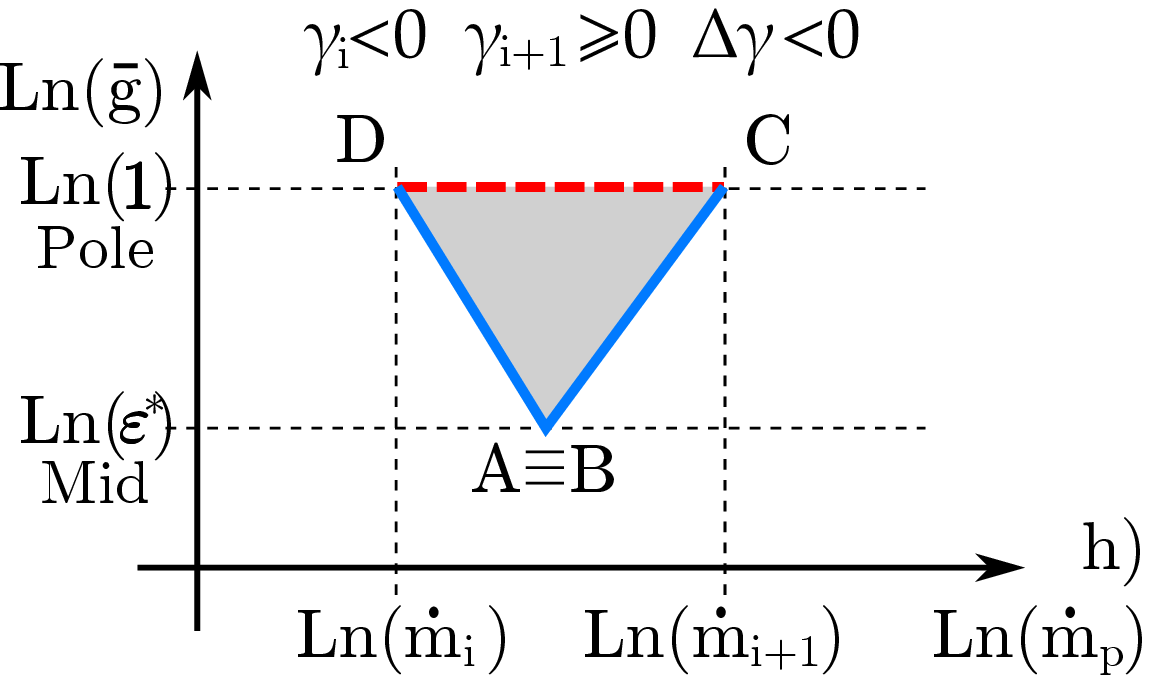}{0.3}
  \caption{These simplified sketches show the possible configurations
    of the allowed ranges of colatitudes where bursts can take place
    as function of $\mdot$ for a single source and burning regime
    $\ii$. The ranges in colatitude are parametrized by $\xbase$ and
    are shown with grey areas. The red dashed segments indicate the
    ignition colatitudes as a function of $\mdot$ when $\beti > 0$,
    the solid blue segments indicate the ignition colatitudes when
    $\beti < 0$. In the case of $\beti = 0$ any colatitude in the grey
    areas is equally probable. Point \Ap{} is the first $\mdot$ at
    which ignition is possible at the highest colatitude allowed,
    \Bp{} the last one. \Dp{} is the first $\mdot$ at which ignition
    is possible at the pole (the lowest colatitude allowed), \Cp{} the
    last. Segments \segu{} correspond to the limit set by
    \eqr{equ:gi}, while segments \segd{} correspond to
    \eqr{equ:gii}. \uplp{Cases a - d}: configurations for cases
    described by $\delgami < 0 \anda \epsstari \lt \ep \trextra{or}
    \delgami \ge 0$, \eqr{equ:exist1}, where the overall minimum to
    $\xbase$ is $\ep$ (the equator). The differences are set by the
    sign of $\gami$ for \segu{} and $\gamip$ for \segd. In order they
    are: positive (or $0$) -- positive ($0$), positive ($0$) --
    negative, negative -- negative, negative -- positive ($0$). Note
    that the actual slopes are given by $1/\gami$ and $1/\gamip$. In
    these cases there is no implied relation between the magnitude of
    $\gami$ and $\gamip$, apart from the respective signs. For example
    the first plot has \segu{} steeper than \segd, but it could also
    be the contrary. The first three plots could even be triangles,
    with the top segment (the pole) collapsed to a point, but at least
    one $\mdot$ should be at the pole, due to the condition $\rhoi \ge
    1$. \uplp{Cases e - h}: same as cases a - d, but for cases
    described by $\delgami < 0 \anda \epsstari \ge \ep$,
    \eqr{equ:exist2}, where the overall minimum to $\xbase$ is $\xsi$
    and the points \Ap{} and \Bp{} coincide. Since $\delgami < 0$,
    $\gamip > \gami$. Under this condition, the second case is
    impossible.}
  \label{fig:TOS}
\end{figure*}

\begin{figure*}
  \centering
  \twofiginarow{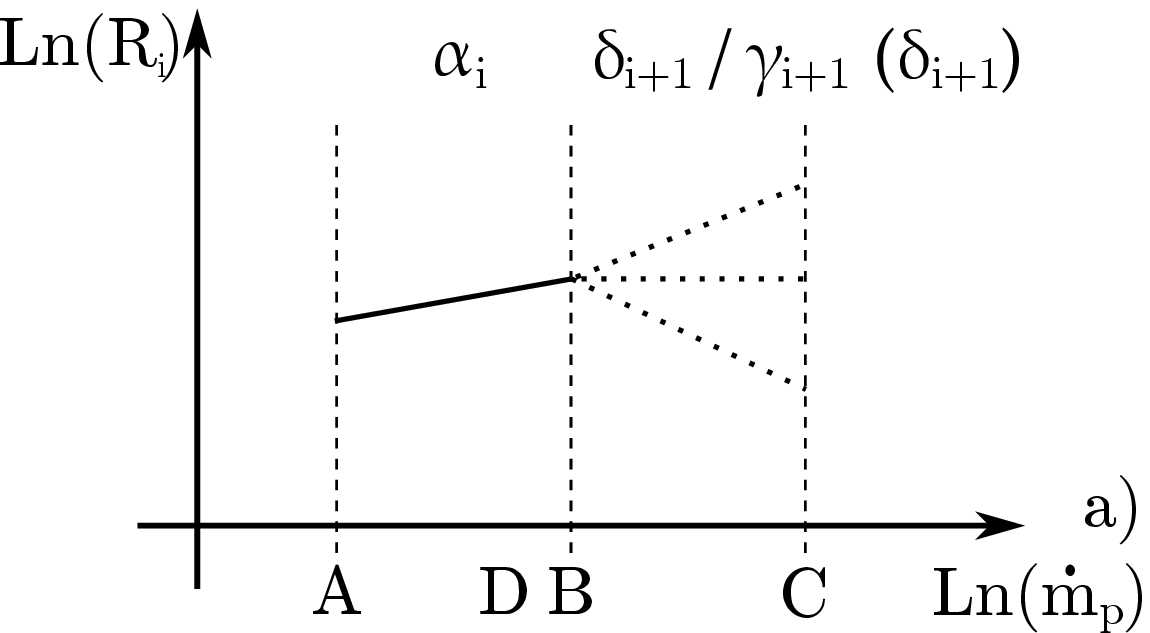}{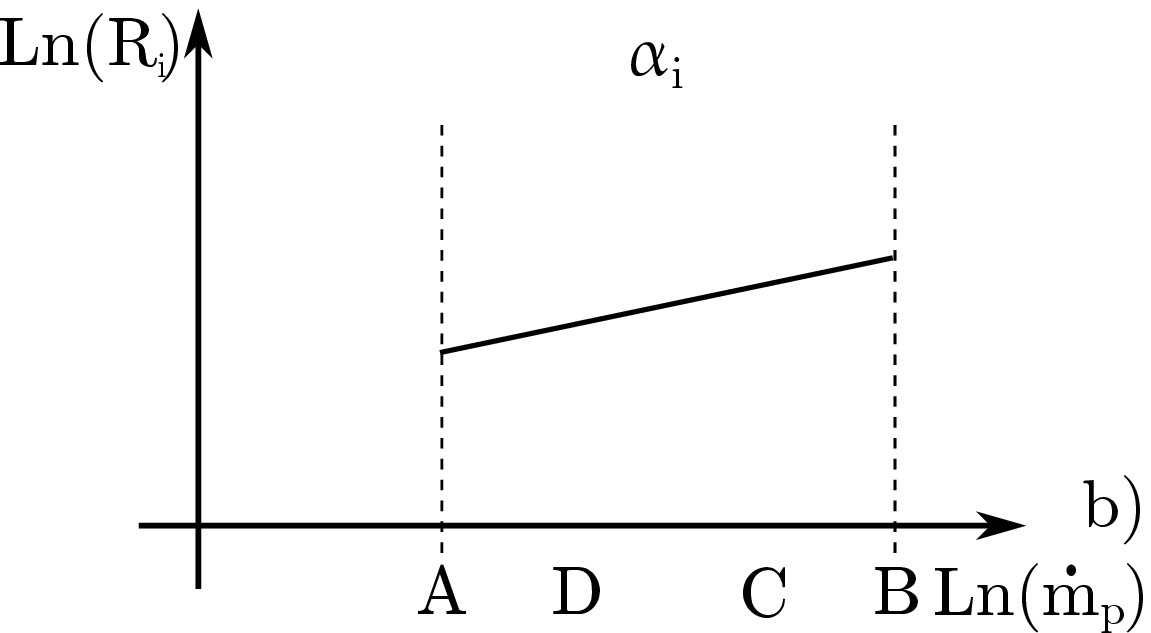}{0.3}\\\vspace{\baselineskip}
  \twofiginarow{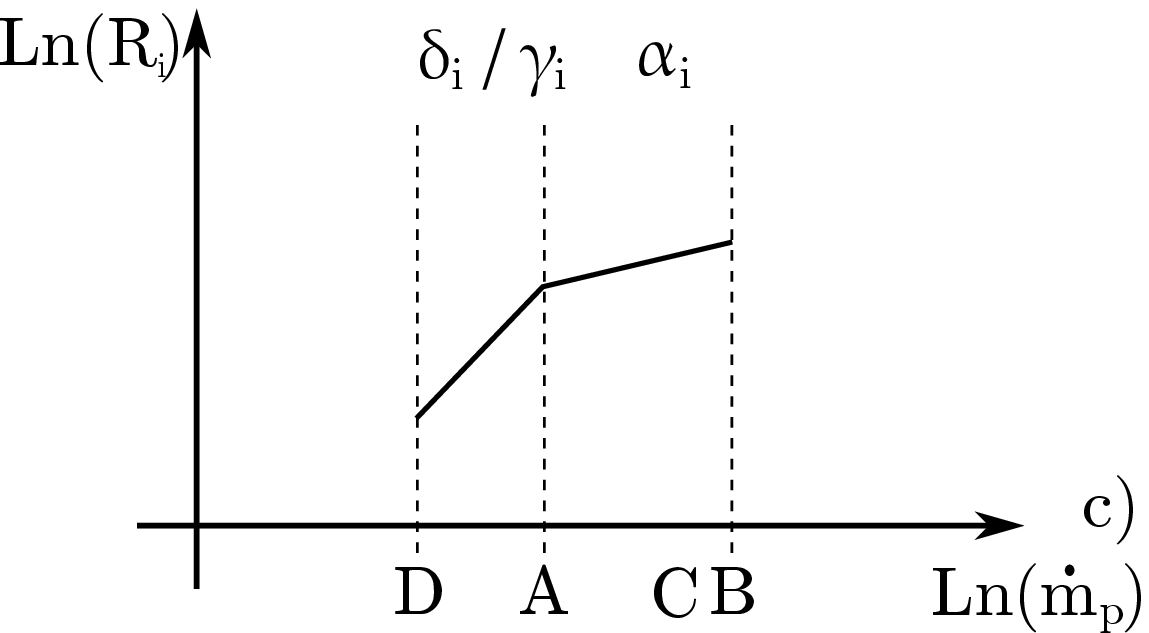}{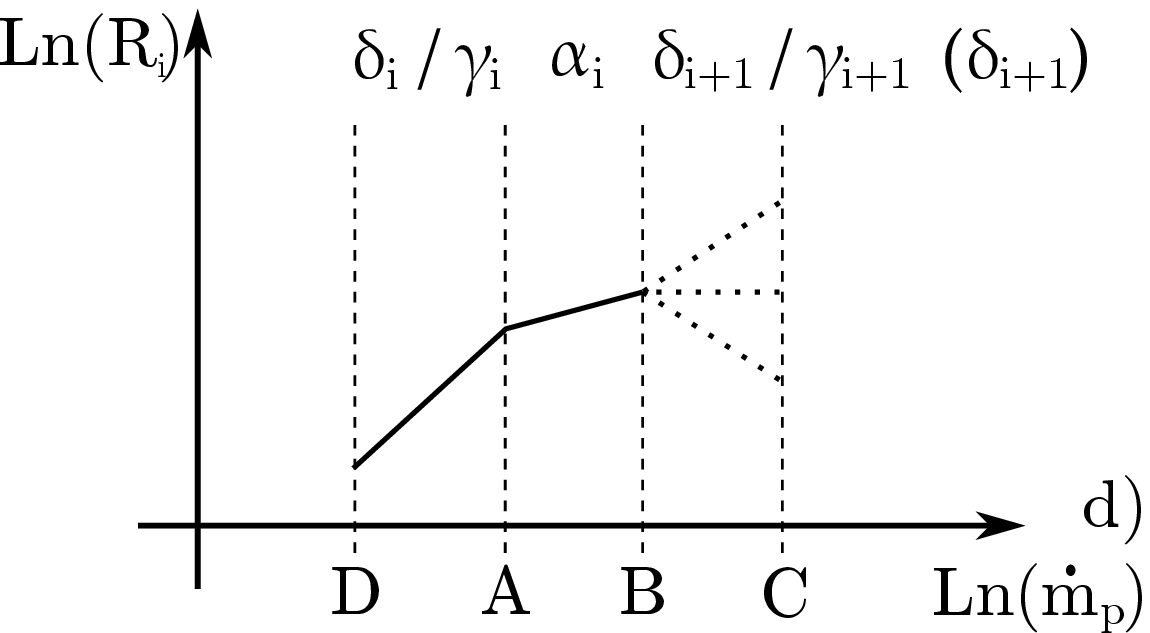}{0.3}\\\vspace{\baselineskip}
  \twofiginarow{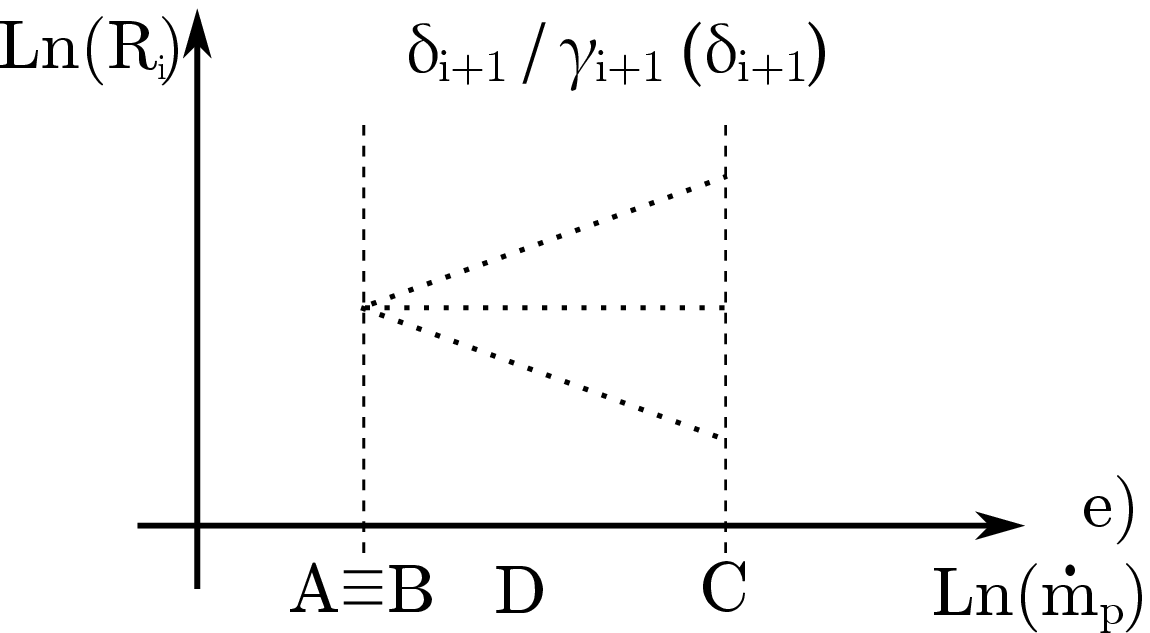}{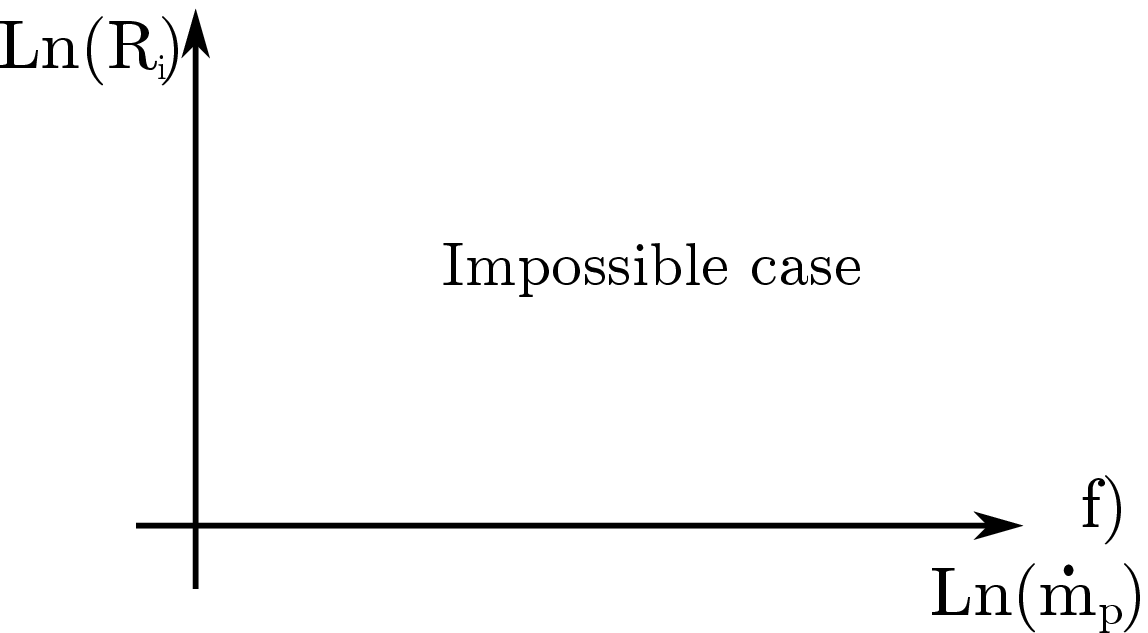}{0.3}\\\vspace{\baselineskip}
  \twofiginarow{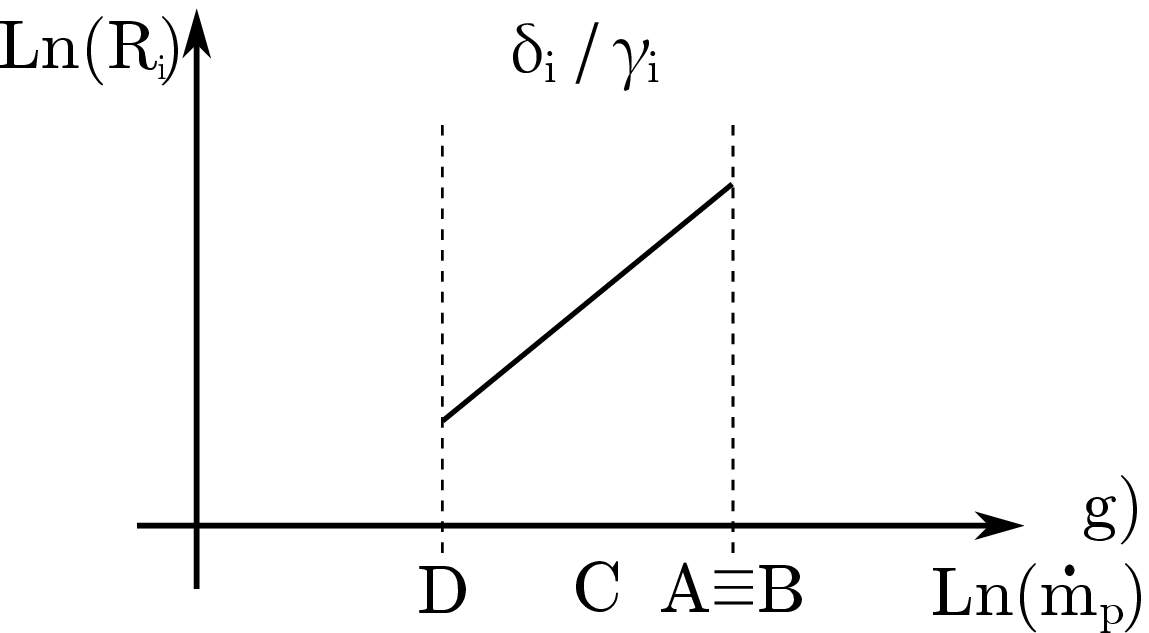}{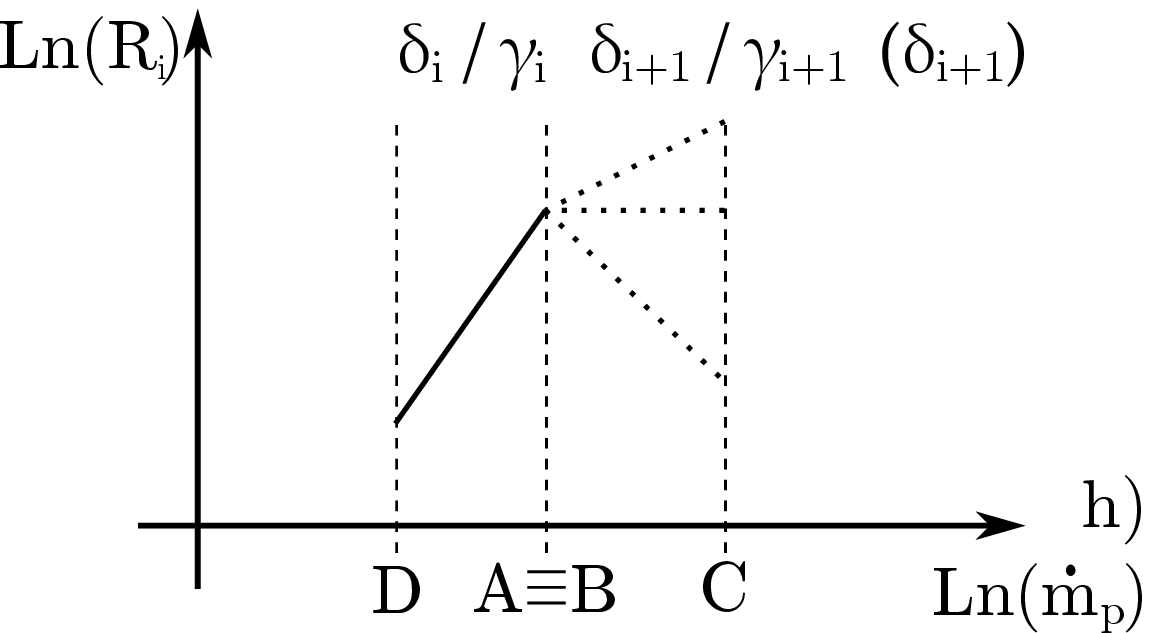}{0.3}
  \caption{Bursting rate evolution of a single source as a function of
    $\mdot$, cases when $\beti < 0$ and $\alpi>0$: these correspond to
    the blue solid \track{s} in \figref{fig:TOS}. The plots shown are
    in one to one correspondence to the plots of \figref{fig:TOS} and
    so are the indicated points \Ap, \Bp, \Cp{} and \Dp{}. Indicated
    above each interval is the slope of the bursting rate. For the
    cases where the slope is $\deliip/\gamip$, the sign of the slope
    is not determined and we show the three possible cases ($> = < 0$)
    using dotted lines. In parentheses we indicate that the sign of
    the slope is dictated by the sign of $\deliip$. On the other hand,
    it is known that $\delii/\gami > \alpi > 0$ when $\delii/\gami$ is
    the slope of the burst rate.}
  \label{fig:RBneg}
\end{figure*}

In order to determine at which colatitude ignition is to be expected,
we first need to know which is the range of allowed $\theta$.  While
\eqrs{equ:exist1} and \eqref{equ:exist2} give the \emph{overall} range
for a given star and burning regime across all possible accretion
rates, the \emph{actual} range at a specific $\mdot$ can be
smaller. The condition that determines this range is given by
\eqr{equ:mdotrange}. Bursts take place only for values of $\mdot$ such
that this relation is satisfied by at least one of the overall allowed
colatitudes.

Then, from \eqr{equ:mdotrange} we can define two functions that will
bound the range of available $\xbase$:

\begin{IEEEeqnarray}{lCl}
  \label{equ:gi}
  \xbl  &=& \left(\frac{\mdot}{\mi}\right)^{1/\gami}\\
  \label{equ:gii}
  \xbh  &=& \left(\frac{\mdot}{\mip}\right)^{1/\gamip}
\end{IEEEeqnarray}
However, these functions can return values greater than $1$ or smaller
than $\ep$ (or $\xsi$), thus violating \eqrs{equ:exist1} or
\eqref{equ:exist2}. In general, the right values to consider are

\begin{IEEEeqnarray}{lCl}
  \label{equ:betgi}
  \xbi   &=& \min \left\{ \max\left[\xbl, \ep(\textrm{or }\xsi) \right], 1 \right\} \\
  \label{equ:betgii}
  \xbip  &=& \min \left\{ \max\left[\xbh, \ep(\textrm{or }\xsi) \right], 1 \right\}
\end{IEEEeqnarray}
and the real ranges for the available values of $\xbase$ at a specific
$\mdot$ are given by:

\begin{IEEEeqnarray}{lCl}
  \label{equ:gmi}
  \xmin  &=& \min \left(\xbi, \xbip \right) \\
  \label{equ:gma}
  \xmax  &=& \max \left(\xbi, \xbip \right)
\end{IEEEeqnarray}
\figref{fig:TOS} shows schematically the various \emph{configurations}
of the available ranges (grey areas) that can be found depending on
the signs of $\gami$ and $\gamip$. Note that $\xbase = 1$ is the pole,
$\xbase = \ep$ is the equator and $\xbase = \xsi$ is somewhere in
between. In the figure, the points \Ap{} and \Bp{} are given by
(see \equlab{s.} \ref{equ:gi} and \ref{equ:gii}):
\begin{IEEEeqnarray}{lCll}
  \label{equ:poiA}
  \ln \mdotgen_{\rm{p, A}} = \ln \mi  &+& \gami & \ln \ep \,(\textrm{or }\xsi)\\
  \label{equ:poiB}
  \ln \mdotgen_{\rm{p, B}} = \ln \mip &+& \gamip& \ln \ep \,(\textrm{or }\xsi)
\end{IEEEeqnarray}
It is clear from the definition of $\xsi$, \eqr{equ:epsstar}, that in
the case of \eqr{equ:exist2} the points \Ap{} and \Bp{}
coincide. Points \Cp{} and \Dp{} correspond, respectively, to
\begin{IEEEeqnarray}{l}
  \label{equ:poiC}
  \ln \mdotgen_{\rm{p, C}} = \ln \mip\\
  \label{equ:poiD}
  \ln \mdotgen_{\rm{p, D}} = \ln \mi
\end{IEEEeqnarray}

Finally, from \eqr{equ:buratebet} it is immediately seen that
$\max\left(\Ri\right) = \Rb\, \mdot^{\alpi} \, \max \left( \x{\beti}
\right)$, so that the answer to the question where does ignition take
place, given a specific $\mdot$? is:

\begin{IEEEeqnarray}{l?l}
  \label{equ:betmag0}
  \beti > 0 \spsp{}
                    & \xbase = \xmax \\
  \label{equ:beteq0}
  \beti = 0 \spsp{} & \forall \xbase \\
  \label{equ:betmin0}
  \beti < 0 \spsp{}
                    & \xbase = \xmin
\end{IEEEeqnarray}
In \figref{fig:TOS}, the ignition colatitudes for the case $\beti > 0$
are shown by the red dashed segments, while for $\beti < 0$ the
colatitudes are indicated by the solid blue segments. Basically $\beti
> 0$ traces the upper boundary and $\beti < 0$ the lower boundary of
the allowed colatitudes. If $\beti = 0$ any colatitude in the grey
areas is equally probable.

\subsection{The bursting rate evolution for a single source}
\label{sec:Revolutionsinglesource}

From an observational point of view, it is interesting to have an idea
of how the bursting rate would evolve within the allowed range of
$\mdot$ depending on the parameters $\alpi$, $\beti$ and $\gami$ and
$\gamip$. In order to study the burst rate evolution for a single
source the starting equation is once again \eqr{equ:buratebet}. In
this section we restrict ourselves to the more physical condition
$\alpi > 0$, the other cases, being an easy extension of these
calculations, are reported in \appref{sec:extra}.

If $\beti = 0$, the bursting rate always grows as

\begin{equation}
  \Ri = \Rbi \mdot^{\alpi}
\end{equation}
and there is not much else to say. The behaviour is more diverse when
$\beti \neq 0$ and requires a more detailed analysis. This is simple
now that we know the \emph{\track}s that ignition colatitude follows
on the $\xbase$ -- $\mdot$ plane, \figref{fig:TOS}. First step is to
know the bursting rate $\Ri$ as a function of $\mdot$ on the various
segments of the plots, then we can combine the different trends
depending on which \track{} is taken. The bursting rates are (using
\equlab{s.} \ref{equ:gi} and \ref{equ:gii}):

\begin{IEEEeqnarray}{lll?l}
  \label{equ:buh1}
  \Ri &= \frac{\Rbi}{\mi^{\beti/\gami}}  & \mdot^{\frac{\delii}{\gami}}   & \textrm{on $\overline{AD}$} \\
  \label{equ:buh2}
  \Ri &= \frac{\Rbi}{\mi^{\beti/\gamip}} & \mdot^{\frac{\deliip}{\gamip}} & \textrm{on $\overline{BC}$} \\
  \label{equ:buh3}
  \Ri &= \Rbi                         & \mdot^{\alpi}              & \textrm{on $\overline{DC}$} \\
  \label{equ:buh4}
  \Ri &= \Rbi\ep^{\beti}                & \mdot^{\alpi}              & \textrm{on $\overline{AB}$, when \eqr{equ:exist1} holds}
 \end{IEEEeqnarray}

It is seen from \eqrs{equ:buh1} and \eqref{equ:buh2} that when
ignition is moving between the pole and the equator (or the maximum
colatitude allowed $\thesi$) the trend is set by the sign and
magnitude of the ratios $\delii/\gami$ and $\deliip/\gamip$. In the
spirit of \figref{fig:TOS} we will not be concerned with the magnitude
of these ratios, which can be determined by numerical simulations or
fitted from observations, but we will study their sign.

\begin{figure*}
  \centering
  \includegraphics[width=0.9\textwidth]{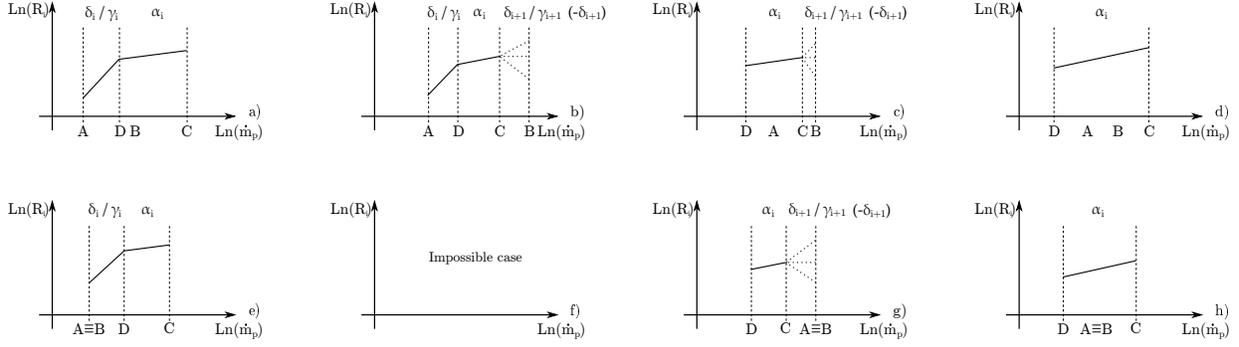}
  \caption{Same as \figref{fig:RBneg}, but for cases when $\beti > 0$
    and $\alpi>0$, red dashed \track{s} of \figref{fig:TOS}. For the
    cases where the slope is $\deliip/\gamip$, it is indicated in
    parenthesis that the sign of the slope is dictated by
    $-\deliip$. Also here $\delii/\gami > \alpi > 0$ when
    $\delii/\gami$ is the slope of the burst rate.}
  \label{fig:RBpos}
\end{figure*}

The expected burst rate evolution for a single burning regime on a
specific source for the cases $\beti \neq 0$ is shown in
\figrefs{fig:RBneg} ($\beti < 0$) and \ref{fig:RBpos} ($\beti > 0$).
One thing to note is that, while the sign of $\delii/\gami$ is not
known in general, in the cases where it is the slope of the function
describing the bursting rate we know it will positive! These cases
are plots c), d), g) and h) of \figref{fig:RBneg} and plots a), b) and
e) of \figref{fig:RBpos}. The sign is known because $\delii/\gami =
\alpi + \beti/\gami$ and for those cases we know that $\beti/\gami >
0$. That also implies that $\delii/\gami > \alpi$, a fact that could
be possibly detected by accurate enough observational campaigns. On
the other hand, the same trick does not apply when we need to know the
sign of $\deliip/\gamip$: plots a), d), e) and h) in
\figref{fig:RBneg} and plots b), c) and g) in \figref{fig:RBpos}.  In
those cases $\beti/\gamip < 0$ and the sign of $\deliip/\gamip$
depends on the difference $\alpi + \beti/\gamip$ or, equivalently, on
the sign of $\deliip$ when $\beti < 0$ (\figref{fig:RBneg}) and the
sign of $-\deliip$ when $\beti > 0$ (\figref{fig:RBpos}).

As an example, we describe now how to obtain plot d) of
\figref{fig:RBneg}. We choose this example because it is one of the
most complicated ones, not because we think this is a more likely
one. For this we need to follow the blue solid line in the
corresponding plot of \figref{fig:TOS}. Ignition starts at the pole
(on \Dp) and proceeds towards the equator as $\mdot$ increases
(\segu). From \eqr{equ:buh1} we know that the rate is increasing
$\propto \mdot^{\delii/\gami}$. When ignition takes place on the
equator ($\overline{AB}$), the bursting rate keeps increasing as
$\mdot^{\alpi}$, but with a lower slope, \eqr{equ:buh4}, since
$\delii/\gami > \alpi$. Finally, for higher $\mdot$, ignition moves
again towards the pole (\segd), and the bursting rate becomes $\propto
\deliip/\gamip$, \eqr{equ:buh2}. It is impossible, on general grounds,
to say if the rate will increase, remain constant or decrease: this
depends on the sign of $\deliip$. The other plots are obtained in the
same way. For example, plot a) is very similar, only the segment
\segu{} is absent; in plot e) also the segment \segt{} is missing,
since ignition starts off equator. The case of plot b) from
\figref{fig:RBpos} is very close in nature to the case of plot d) of
\figref{fig:RBneg}, however reversed. Here ignition is initially on
the equator, \Ap{,} and then moves towards the pole on \segu{}
(following the red path in \figref{fig:TOS}), the burst rate growing
as $\mdot^{\delii/\gami}$. While the flame ignites preferentially at
the pole, on \segq{,} the rate grows as $\mdot^{\alp}$ since the
normalization factor due to $\xbase$ stays constant. Finally, after
the point \Cp{} as been reached, ignition moves again towards the
equator with the burst rate evolving as $\mdot^{\deliip/\gamip}$. If
$\deliip/\gamip < 0$ the burst rate will be observed to decrease.
However, very differently from the cases when $\bet < 0$, the
accretion rate at which the burst rate is seen to peak is constant:
$\mip$. Both in \figrefs{fig:RBneg} and \ref{fig:RBpos} a negative
$\deliip/\gamip$ would result in the burst rate starting to decrease
after some value of $\mdot$ ($\mdott$).

\section{Example application}
\label{sec:exabild}

Here we provide an explicit example of the formalism of this paper,
showing how the results for helium burning of \citet{rev-1998-bild}
and \citet{art-2007-piro-bild} translate into the parameters $\alp$,
$\bet$ etc. We start with the case of gravity, which is the most
developed. In doing so, we repeat some of the formulae from the
author\footnote{The formulae will look slightly different because we
  have rederived them in order to keep explicit all the terms that
  involve the composition, we avoided rounding numbers in intermediate
  steps and we apply no scaling to variables like $\mdotgen$ or
  $\gfe$. We keep the opacity $\kappa$ explicitly instead of inserting
  the electron scattering formula $\kappa_{\rm{es}} =
  \sigma_{\rm{Th}}(1 + X) / (2 m_{\rm{p}})$.}. This is also the regime
initially described by \citet{art-2007-coop-nara-a}.

First we show how to obtain the bursting rate parameters (see
\equlab{s.}  \ref{equ:trec} - \ref{equ:buratebet}). In the case of
ignition in a pure helium environment, the ignition depth is given by
\alienequ{20} of \citet{rev-1998-bild}:

\begin{equation}
  \label{equ:ypureHe}
  \yi{1} = 1.08\tent{14} (Y\mu E_{18} \kappa)^{-2/5} \mdotgen^{-1/5} \gef{-2/5}
  \gcm
\end{equation}
where $Y$ is the helium mass fraction, $E_{18}$ is the energy released
per unit mass by the burning in units of $10^{18}$ erg g$^{-1}$ and
$\kappa$ is the opacity in cm$^2$ g$^{-1}$.

$\R = \locmdot / y_{\rm{ign}}$, therefore

\begin{equation}
  \label{equ:bildpurehe}
  \R_1 = 9.28\tent{-15} (Y\mu E_{18} \kappa)^{2/5} \mdotgen^{6/5} \gef{2/5} \Hz
\end{equation}

Comparing this to \eqrs{equ:protorate} and \eqref{equ:buratebad}, and
expanding $\gfe = \gep \xbase$, it is seen that $\alptn{1} = 6/5$,
$\bettn{1}=2/5$, so that
\begin{equation}
  \label{equ:abHe}
  \alpn{1} = 6 / 5 \;\textrm{and}\; \betn{1} = -4 / 5
\end{equation}
The pseudo constant $\Rb_1 = 9.28\tent{-15} (Y\mu E_{18} \kappa)^{2/5}
\gep^{-4/5}$ Hz (g s$^{-1}$ cm$^{-2}$)$^{-6/5}$. It is evident how the
properties of composition, opacity, burning regime and stellar mass
and radius are contained in $\Rb$.

When helium burns in a mixed hydrogen--helium environment and flux
from the bottom can be neglected, \alienequ{32} of
\citet{rev-1998-bild}, the ignition depth is:

\begin{equation}
  \yi{2} = 2.55\tent{10} Y^{-1/3} \zcn^{-5/18} \mu^{-2/9}  \kappa^{-7/18} \gef{-2/9}
  \gcm
\end{equation}
here $\zcn$ is the metallicity, i.e. mass fraction of carbon, nitrogen
and oxygen. Note that $\yi{2}$ is independent of $\mdotgen$, even
though this is not always the case \citep[at high accretion rate
  and/or low metallicity][\alienequlab{.}
  37]{rev-1998-bild}. Therefore

\begin{equation}
  \label{equ:bilmixedhe}
  \R_2 = 3.92\tent{-11} Y^{1/3} \zcn^{5/18} \mu^{2/9}  \kappa^{7/18} \mdotgen{}  \gef{2/9} \Hz
\end{equation}
so that $\alptn{2} = 1$, $\bettn{2}=2/9$; then
\begin{equation}
  \label{equ:abmixedHe}
  \alpn{2} = 1 \;\textrm{and}\; \betn{2} = -7 / 9
\end{equation}
Furthermore, $\Rb_2 = 3.92\tent{-11} Y^{1/3} \zcn^{5/18} \mu^{2/9}
\kappa^{7/18} \gep^{2/9}$ Hz (g s$^{-1}$ cm$^{-2}$)$^{-1}$.

Second, we provide examples for the limits in mass accretion rate,
\eqrs{equ:mdotlimbad} and \eqref{equ:mdotlimbet}, for the validity of
the bursting rate of each of these burning regimes. In the case of
pure helium bursts, the lower limit is set by the stability of the
hydrogen burning, \alienequ{36} of \citet{rev-1998-bild}, which
otherwise would be bursting before helium could:

\begin{equation}
  \label{equ:mdotHElow}
  \mdotgen_{1,\low} = 4.18\tent{-3} X^{-1} \zcn^{1/2} \kappa^{-1/2}
  \gscm
\end{equation}
independent from gravity. The upper limit is set by the requirement
that helium ignites at a depth where all hydrogen is depleted,
\alienequ{35} of \citet{rev-1998-bild}:

\begin{equation}
  \label{equ:mdotHEhigh}
  \mdotgen_{1,\hig} = 2.32\tent{2} \zcn^{13/18} X^{-1} Y^{-1/3} \mu^{-2/9} \kappa^{-7/18} \gef{-2/9}
  \gscm
\end{equation}
This means that $\mdotgen_{1} = 4.18\tent{-3} X^{-1} \zcn^{1/2}
\kappa^{-1}$ and $\mdotgen_{2} = 2.32\tent{2} \zcn^{13/18} X^{-1}
Y^{-1/3} \mu^{-2/9} \kappa^{-7/18} \gep^{-2/9}$. Furthermore,
$\gamtn{1} = 0$, $\gamtn{2}=-2/9$ and so
\begin{equation}
\gamn{1} = 1 \;\textrm{and}\; \gamn{2} = 7 / 9
\end{equation}
Combining these with \eqr{equ:abHe}, we have for \eqrs{equ:delii},
\eqref{equ:deliip} and \eqref{equ:delgami}

\begin{align}
  \delnn{1}{1} &= 2/5  \\
  \delnn{1}{2} &= 2/15 \\
  \delgamn{1}  &= 2/9
\end{align}

For the case of helium ignition in a mixed hydrogen--helium
environment, the lower limit is set by the upper limit of pure helium
ignition, $\mdotgen_{2,\low} = \mdotgen_{1,\hig}$.  The upper limit is
set by the stability of helium burning in this mixed composition
conditions, \alienequ{24} of \citet{rev-1998-bild}. This is:

\begin{equation}
  \label{equ:mdotHHEhigh}
  \mdotgen_{2,\hig} = 1.79\tent{-7} Y^{1/2} \mu^{1/2} E_{18}^{-3/4} \kappa^{-3/4} \gef{1/2}
  \gscm
\end{equation}
which leads to $\mdotgen{_3} = 1.79\tent{-7} Y^{1/2} \mu^{1/2} E_{18}^{-3/4}
\kappa^{-3/4} \gep^{1/2}$, $\gamtn{3} = 1/2$ and
\begin{equation}
\gamn{3} = 3 / 2
\end{equation}
This implies, with \eqr{equ:abmixedHe},

\begin{align}
  \delnn{2}{2} &= 0      \\
  \delnn{2}{3} &= 13/18  \\
  \delgamn{2}  &= -13/18 
\end{align}
For a NS with $M=1.4$ \ms, $\Rs=10$ km, accreting solar composition
$X=0.7$, $Y=0.29$, $\zcn=0.01$ with the opacities reported by
\citet{rev-1998-bild}, we have $\Rb_{1} = 2.08\tent{-8}$ Hz (g
s$^{-1}$ cm$^{-2}$)$^{-6/5}$, $\Rb_{2} = 2.75\tent{-9}$ Hz (g s$^{-1}$
cm$^{-2}$)$^{-1}$, $\mdotgen_{1}=6.69\tent{2} \gscm$,
$\mdotgen_{2}=4.72\tent{3} \gscm$, $\mdotgen_{3}=1.33\tent{5} \gscm$
(this value is actually $\sim 1.5$ times the local Eddington limit
$\mdotedd = 2 c m_{\rm{p}} / [\sigma_{\rm{Th}}(1 + X)] = 8.88\tent{4}$
g s$^{-1}$ cm$^{-2}$)\footnote{This case is interesting because it
  shows that $\rhoi > 1$ even though the numerical coefficient of
  \eqr{equ:mdotHHEhigh} is smaller than the one of
  \eqr{equ:mdotHEhigh}.} and we have

\begin{IEEEeqnarray*}{rlll}
                     \R_1             &= 2.08\tent{-8}& \mdotgen^{6/5}          \x{-4/5}  &\Hz\\
                     \R_2             &= 2.75\tent{-9}& \mdotgen^{\phantom{6/5}}  \x{-7/9}  &\Hz\\
                     \mdotgen_{1,\low} &= 6.69\tent{2}&                         \x{}      &\gscm\\
  \mdotgen_{1,\hig} =  \mdotgen_{2,\low} &= 4.72\tent{3}&                         \x{7/9}   &\gscm\\
                     \mdotgen_{2,\hig} &= 1.33\tent{5}&                         \x{3/2}   &\gscm
\end{IEEEeqnarray*}

For the case of ignition in a pure helium environment we have $\gami =
\gam_1 > 0$, $\gamip = \gam_2 > 0$ and $\delgamn{1} > 0$, which
corresponds to \eqr{equ:exist1} and to plot a) of
\figref{fig:TOS}. $\bet_1<0$, which according to \eqr{equ:betmin0}
means ignition will take place at $\xmin$. Therefore, as $\mdot$
increases, $\xbase_{\rm{ign}}$ will trace the lower boundary of the
grey area (solid blue segments): starting on point \Ap{,} ignition
will be at the equator until the segment \segd{} begins, at which
point ignition will move towards the pole following this segment. The
case of helium ignition in a mixed hydrogen--helium environment is
similar, having $\bet_2<0$, $\gami=\gam_2>0$, $\gamip=\gam_3>0$, but
$\delgamn{2}<0$. In this case the behaviour is different for slow and
fast rotators, where fast means $\ep < \epsstar_2 = ( 1.33\tent{5} /
4.72\tent{3})^{1 / -(13/18)} = 9.83\tent{-3}$ or equivalently $\nu >
\nuk \sqrt{1 - \epsstar_2} = 9.95\tent{-1}\nuk$. For slow
rotators\footnote{Note that in this case almost every NS would be a
  slow rotator, since the limit is very close to the mass shedding
  limit.} the evolution is again described by the lower boundary of
plot a) of \figref{fig:TOS}, but for fast rotators the available
ignition colatitudes are described by \eqr{equ:exist2} and plot e) of
\figref{fig:TOS}. For fast rotators ignition begins off equator (at
$\theta^*_{2}$, $\xbase = \epsstar_2$) on point $\rm{A}\equiv\rm{B}$
and moves polewards along the segment \segd.

Since $\bet_1 < 0$ and $\bet_2 < 0$, the bursting rate evolution is
described by the plots a) and e) of \figref{fig:RBneg}. $\delnn{1}{2}
> 0$, so that plot a) tells us that we would expect an always growing
bursting rate with increasing $\mdot$ for pure helium burning, with a
change of slope at some point. Since also $\delnn{2}{3} > 0$, plots a)
and e) predict the same for bursts of helium ignition in a mixed
hydrogen and helium environment, with the faster sources displaying
one single slope. Since the maximum burst rate is attained at the
pole, it is independent of the rotation of the star and so is the mass
accretion rate of the peak, \eqrs{equ:poiC} and \eqref{equ:buh3}.

We now move on to see how the results of \citet{art-2007-piro-bild}
translate into our formalism. The main point to make is that the
powers in the formulae of those authors should change signs, since we
suggest to have $\xbasem$ depend on the inverse of $\nu$ in order to
have the minimum of $\xbasem$ at the equator. As for the burst rate,
\alienequ{70} of \citet{art-2007-piro-bild} would read

\begin{equation}
  \R \propto \locmdot^{1.25}\xm{0.36}
\end{equation}
so that $\alp = 1.25$ and $\betm = 0.36$. The authors also report two
limits for their regime of mixing modified helium burning:

\begin{align}
  \mlt \propto& \xm{-3}\\
  \mht \propto& \xm{-0.62}
\end{align}
so that $\gamml=-3$ and $\gammh=-0.62$. Note that also in the case of
mixing the analytical predictions would give a consistently increasing
burst rate. $\betm>0$ and both $\gammg{*}<0$, so that the case is that
described by plots c) or g) of \figrefs{fig:TOS} and \ref{fig:RBpos}
($\delgami < 0$). These cases allow for decreasing burst rate, but
here $\deliip/\gamip = 0.67 > 0$: the expected rate is
increasing. However, once again, these are simplified analytical
calculations and some differences with real burst physics are to be
expected \citep[see e.g.][who include a more elaborated version of the
  Tayler-Spruit dynamo and also find that at high spin hydrodynamical
  instabilities become efficient]{art-2009-kee-lang-zand}.
  
It is curious to note how both the case of \citet{rev-1998-bild} and
the case of \citet{art-2007-piro-bild} do actually fall in the
categories that would give decreasing burst rate \emph{if the ratio
  $\deliip/\gamip$ were negative.} The values of $\alp$, $\bet$ and
$\gam_*$ are uncertain enough that this could be happening in
actuality. Between the two mechanisms mentioned above, we think mixing
is the best candidate.

\section{Summary and discussion}
\label{sec:discussion}

\subsection{The role of local conditions}

We have presented simple analytical relations that would enable a
comparison between models and observations. In
\secref{sec:definitions} we began introducing the relation between the
observed total mass accretion rate $\mdott$ (as measured near the
star, via the average local accretion rate $\mdotg$, $\mdott = 4 \pi
\Rs^2 \mdotg$) and the local $\mdot$ at the pole in
\eqr{equ:mpmg}. This relation is used to facilitate the calculations
since it allows us to compare one single observational piece of
information, $\mdott$, to one single theoretical piece of information,
$\mdot$. However, as we noted, even up to $\nu=10^3$ Hz \citep[the
  fastest known NS spins at $716$ Hz][and the fastest burster spins at
  $620$ Hz \citealt{art-2002-muno-etal}]{art-2006-hessels-j-etal} the
difference between $\mdotg$ and $\mdot$ is just of order $10$\%. Then,
in \secrefs{sec:definitions} and \ref{sec:ignitionmp} we generalized
the work of \citet{art-2007-coop-nara-a} and presented a description
of the burst rate $\R$ vs $\mdot$. We parametrized the burning physics
with various parameters ($\xbase$, $\alpi$, $\beti$, $\gam_*$ and
$\mdotgen_*$). $\xbase$ is a function of the colatitude $\theta$ and
the spin frequency $\nu$. It is set by the dependence of the burning
physics on local conditions. We discussed two possible mechanisms that
may have an effect: local gravity, as explored by
\citet{art-2007-coop-nara-a}, and mixing, as explored by
\citet{art-2007-piro-bild}. The two mechanisms establish different
relations between $\nu$, $\theta$ and the burning physics, which are
summarized by \norm{.} In the case of the effective gravity, $\xbase$
is the ratio $\gef{} / \gep$, \eqrs{equ:xbaseeff} and
\eqref{equ:xbasedef}. In the case of mixing this dependence has not
been worked out in full form yet \citep[but see][]{art-1999-spruit,
  art-2002-spruit, art-2007-piro-bild} and we just hint at a
possibility in \eqr{equ:xbasem}.

The form of $\xbase$ is very important because we use it to express
the colatitude of ignition once $\nu$ is fixed. While $\alpi$
expresses the dependence of the burst rate on the mass accretion
$\mdot$, $\beti$ expresses the importance of each of the mechanisms
that are at work in setting the burst rate, \eqr{equ:buratebet}. In
the case of the changes to local gravity due to the centrifugal force,
$\bet$ is determined by the dependence of the ignition depth and the
temperature profile of the column on gravity \citep{rev-1998-bild}. In
the case of mixing, it is determined by the dependence of those very
same quantities on the rotation-shear induced mixing
\citep{art-2007-piro-bild}. The $\mdotgen_*$ are the boundaries of
accretion rate where bursts can take place as calculated in the
absence of rotation, e.g. at the pole; these boundaries at other
colatitudes depend also on the local conditions via $\x{\gam_*}$,
\eqr{equ:mdotlimbet}. As it can be seen, the local conditions, be they
set by the effective gravity, mixing or other mechanisms, are very
important, because they control very strongly the evolution of the
burst rate. In \secref{sec:ignitionmp} we discussed the case $\alpi >
0$, i.e. when locally the burst rate increases with accretion rate. We
provided summarizing formulae and plots for the ignition colatitude
and burst rate as a function of $\mdot$ ($\mdott$): \eqrs{equ:betmag0}
- \eqref{equ:buh4} and \figrefs{fig:TOS} - \ref{fig:RBpos}. In the
Appendices we provide similar results for other, less likely cases.

Due to their nature, the equations were derived under somewhat
simplified assumptions, which could be improved. First, general
relativistic corrections to $\gef{}$ could be taken into
account. \citet{art-2014-alge-mor} for example show that rotation
introduces further terms to the ratio $\gfe{} / \gep$ that we do not
include in \eqr{equ:xbasedef}. These terms depend on the oblateness of
the star and the mass quadrupole moment. These corrections have a
different form from \eqr{equ:xbasedef} and can be higher even for
stars rotating at $500$ Hz \citep{art-2014-alge-mor}. Thus, they would
change some of the quantitative conclusions drawn from the equations
of this paper. The nature of the conclusions should not be
affected. Second, the local accretion rate depends on $\nu$ and
$\theta$ only through the effective gravity term,
\eqr{equ:mdrel}. This may not be the case depending on the extent of
the boundary layer \citep{rev-2000-bild} or if some form of
confinement is operating, for example due to magnetic fields. This may
change \eqr{equ:mdrel} and therefore most of the following
equations. Third, there may even be a dependence on $\mdott$ of the
extent in $\theta$ of the boundary layer or of the size of the
accretion column in case of strong magnetic fields: this would even
make \eqrs{equ:mdrel} and \eqref{equ:mpmg} non linear in
$\mdott$. Finally, extra heating in the upper layer where accretion
takes place may affect ignition depth, burst rate and the boundaries
in mass accretion rate as in \eqrs{equ:buratebet} and
\eqref{equ:mdotlimbet}. This effect could arise from a magnetic
hot-spot or if some heating mechanism is at work at the accretion disc
boundary layer \citep[as suggested by][]{art-1999-ino-suny,
  art-2010-inoga-suny}. These effects would introduce different
dependencies on the spatial position ($\theta$, $\phi$) and, in the
case of the boundary layer, also on $\nu$, therefore $\xbase$ would
have a different form. Including these dependencies may contribute to
further refine the equations here presented. We leave this for future
work.

We continued in \secref{sec:exabild} presenting an example application
which shows how the equations and plots of \secref{sec:ignitionmp} and
the appendices could be used, after the $\nu$-dependent conversion
between $\mdott$ and $\mdot$ has been applied. We showed in a
straightforward way that the dependencies predicted by theory (in this
case the values of $\alpi$, $\beti$, $\gam_*$ and $\mdotgen_*$ based
on the simplified analytical calculations of \citealt{rev-1998-bild}
and \citealt{art-2007-piro-bild}) would not agree with observations,
since they predict consistently increasing burst rate vs $\mdott$,
even taking into account the effects of local gravity and mixing.

\subsection{A mechanism for decreasing burst rate}

The second goal (and a very exciting conclusion) of this paper is a
possible explanation that naturally accounts for two observational
oddities: decreasing burst rate with increasing $\mdott$, and the
weakness of the high $\mdott$ bursts. The decrease in burst rate after
a certain accretion rate $\mdott$ is relatively common \citep[see
  e.g.][]{art-2003-corne-etal}. The reason behind this decrease has
been a mystery for many years. It has been explained either as a
consequence of a switch to a burning regime with intrinsically
decreasing burst rate, $\alpha < 0$ in our formalism
\citep{art-2003-nara-heyl, art-2007-coop-nara-a} or as a change in
accretion geometry that changes the local $\mdotgen$
\citep[see][]{rev-2000-bild, rev-2003-2006-stro-bild-book}. We think
we can explain it with the effect that the local conditions have on
the burst rate, e.g. due to effective gravity or mixing. From the
plots in \secref{sec:ignitionmp}, \figrefs{fig:RBneg} and
\ref{fig:RBpos}, it can be seen that it is actually possible to have a
decreasing burst rate, \emph{even if $\alpi>0$ locally all the
  time}. The condition for this is that $\deliip/\gamip = \alpi +
\beti / \gamip < 0$. The physical meaning of this combination is as
follows.

\begin{figure*}
  \centering
  \includegraphics[width=0.43\textwidth]{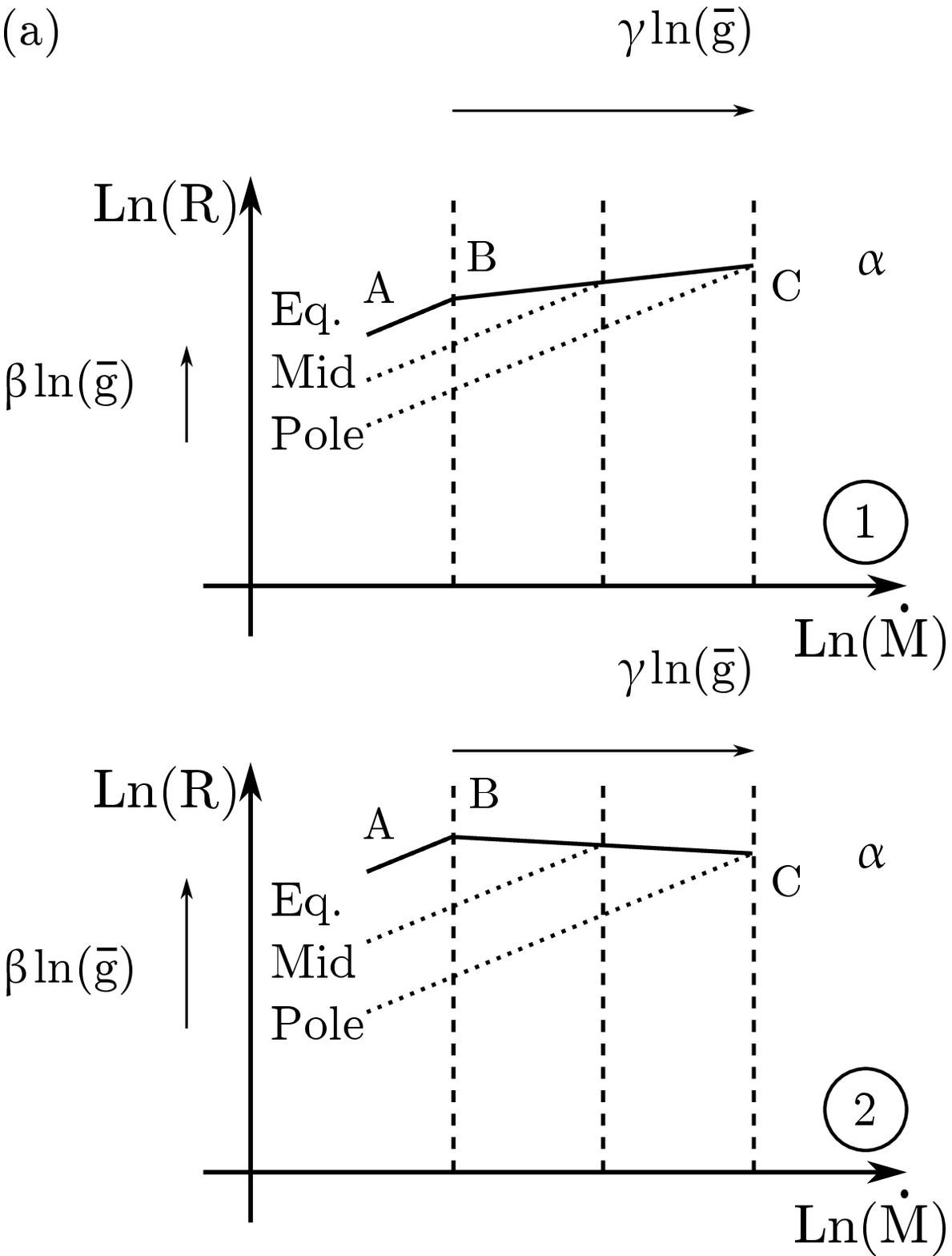}
  \hspace{\stretch{1}}
  \includegraphics[width=0.43\textwidth]{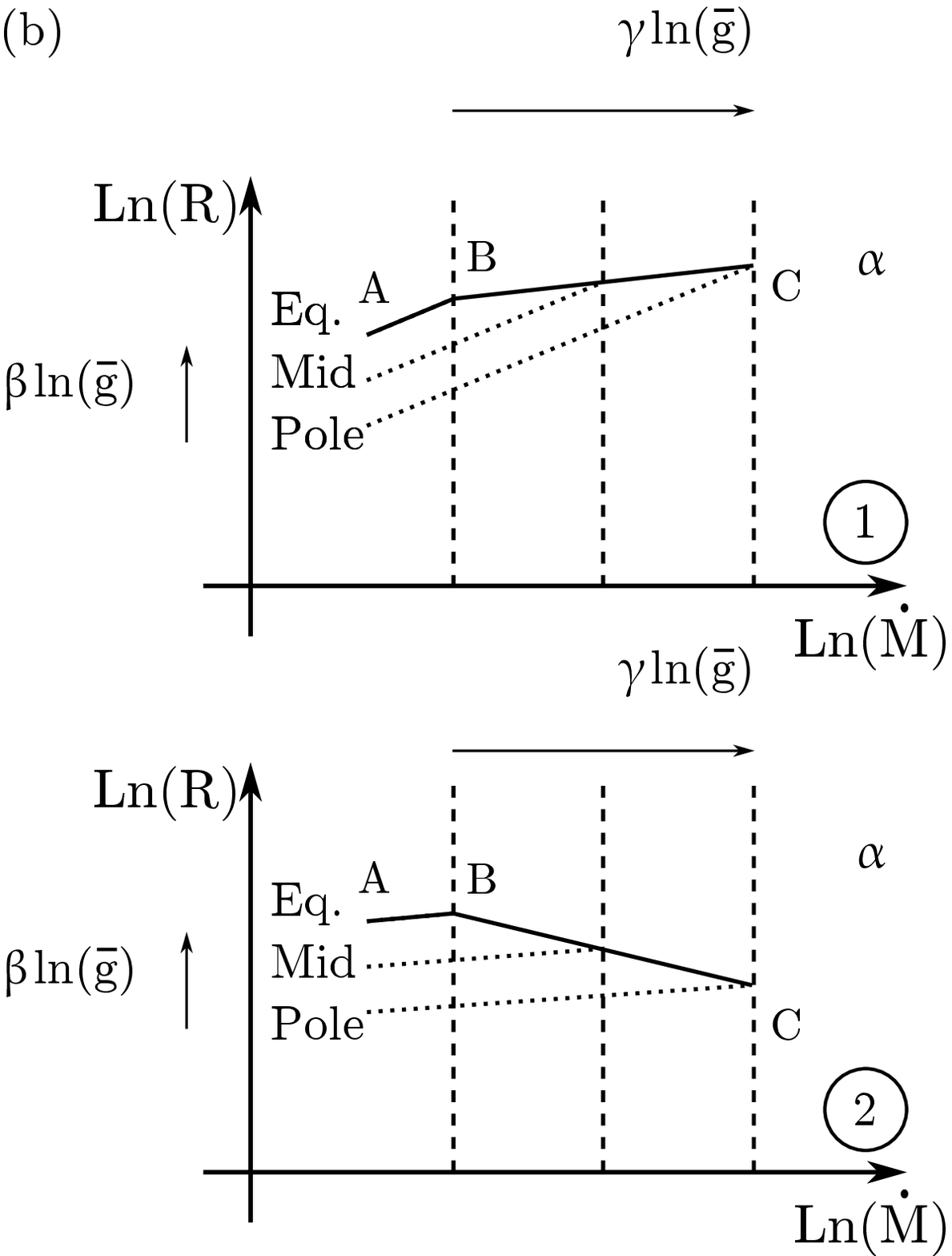}
  \caption{Same as \figref{fig:sketch}, but highlighting the effects
    of $\bet$, panel (a), and $\alp$, panel (b). \uplp{Panel (a)}. At
    fixed $\alp$ and $\gamip$, if $\bet<0$ and $|\bet|$ is small, Case
    $1$, then the increase in burst rate due to the increasing \inc{}
    can cover the gap due to the normalization factor $\xbase^\bet$
    and the burst rate is seen to increase. When $|\bet|$ is large,
    Case $2$, the gap is too wide and the increasing burst rate cannot
    compensate for it: the burst rate is seen to decrease. \uplp{Panel
      (b)}. At fixed $\bet$ and $\gamip$ if $\alp$ is high enough,
    Case $1$, then the burst rate keeps increasing also when ignition
    moves off equator. If $\alp$ is low, Case $2$, then the increasing
    \inc{} cannot compensate for the normalization factor and the
    burst rate is seen to decrease. For both panels in Case $1$
    $\deliip/\gamip >0$ and in Case $2$ $\deliip/\gamip <0$.}
  \label{fig:sketch2}
\end{figure*}

Consider a case with $\beti<0$, $\gami>0$ and $\gamip>0$; we will
highlight the role of each parameter separately, starting with
$\gamip$ (see also \secref{sec:summary} and \figref{fig:sketch}). The
burst rate is given by \eqr{equ:buratebet}, $\R \propto \mdot^{\alpi}
\x{\beti} \propto \mdott^{\alpi} \x{\beti}$. The factor \norm{} sets
the difference between the burst rate at the equator and the pole (and
also all the other colatitudes). Since $\xbase(\theta=\pi/2) <
\xbase(\theta=0)$ and $\beti<0$ the burst rate at the equator is
higher and the bursts initially ignite there. As long as the equator
can burst, the rate in this phase will grow as $\mdott^{\alpi}$. When
the equator stabilizes, the ignition site moves polewards at a
\speed{} $\Delta \ln \xbase / \Delta \ln \mdott = 1 / \gamip$. It will
reach the pole in $\Delta \ln \mdott = -\gamip \ln \ep$, where $\ep$
is $\xbase$ evaluated at the equator. The rate will be $\Rb \propto
\mdott^{\alpi + \beti/\gamip}$. If the ignition moves towards the pole
in a range $\Delta \ln \mdott$ which is wide enough and the growth of
burst rate due to $\mdott^{\alpi}$ is able to compensate the initial
gap due to $\x{\beti}$, then the burst rate will increase (large
$\gamip$, Case $1$ in \figref{fig:sketch}). If the ignition moves
towards the pole in a range $\Delta \ln \mdott$ which is too narrow,
then the increase of burst rate due to $\mdott^{\alpi}$ will not be
able to overcome the initial gap and the burst rate will decrease
(small $\gamip$, Case $2$ in \figref{fig:sketch}). If $\beti>0$ the
situation is analogous, with pole and equator exchanging roles. This
time the pole has the advantage, see for example plot c) in
\figref{fig:RBpos}. When ignition leaves the pole towards the equator
on the segment \segd{,} the growth in burst rate due to
$\mdott^{\alpi}$ can or can not compensate the initial gap due to
$\x{\beti}$ depending on the value of the interval $\Delta \ln \mdott
= \gamip \ln \ep$. Note that in the case $\beti>0$ we need
$\gamip<0$. The burst rate is of course $\propto \mdott^{\alpi +
  \beti/\gamip}$. In \figref{fig:sketch2} we describe how differences
in $\bet$ and $\alp$ can have similar effects. In panel a) we show the
effect of $\bet$. $\bet$ sets the gap between the burst rate at
different colatitudes. If $\bet<0$, the higher $|\bet|$ the higher
this gap will be (since $\xbase\le1$). For high $|\bet|$ the increase
\inc{} will not be able to compensate the gap and burst rate will
decrease, Case $2$. If $|\bet|$ is small enough, the gap can be
covered and the burst rate will increase, Case $1$. In panel b) we
show the role of $\alp$, which is apparent by now. If $\alp$ is high
enough, \inc{} will be high and will be able to cover the gap due to
the normalization factor \norm{,} Case $1$. Otherwise, the rate will
be seen to decrease, Case $2$. The physical meaning of the condition
$\alpi + \beti/\gamip < 0$ is then this: that the resulting rate when
ignition moves off the initially favourite site is a competition
between the increase in rate set by $\alpi$ and the initial gap set by
$\beti$ \emph{compensated} by the \speed{} $\Delta \ln \xbase / \Delta
\ln \mdott$ set by $1/\gamip$.

This simple mechanism can explain quite naturally the decrease in
burst rate with the initial gap in burst rate, the process of
stabilization of the bursts and the migration of the ignition to other
colatitudes. It is also very appealing because it reconciles the
observations with the time dependent 1D simulations that predict
consistently increasing burst rate $\alpi > 0$.  At the same time,
since a smaller fraction of the star is available for the unstable
burning of regime $i$, the rest of the star would be burning
stably. The stable burning would reduce the available fuel for the
spreading flame of the bursts after ignition took place, thus
explaining the other observational feature: less energetic bursts
\citep[and reference therein]{art-1988-par-pen-lew,
  rev-2003-2006-stro-bild-book}. If this scenario is true, it may have
implications also for all the observational attempts at measuring the
NS radius that exploit the bursts, since not all the area of the star
may be emitting, at least not homogeneously. Furthermore, the decrease
in burst rate and the weakening of the bursts would make them more
difficult to detect before the theoretical limit is reached and this,
combined with other stabilizing effects
\citep[e.g.][]{art-2009-kee-lang-zand}, would give the impression of
stabilization before the expected theoretical value of the accretion
rate boundary.

The last sentence needs some refinement. We presented two cases where
it is possible for the burst rate to decrease: $\bet<0$, with bursts
initially igniting on the equator and $\bet>0$ with bursts initially
at the pole. The value of $\Delta\ln\mdott$ between the peak of the
burst rate and the end of the bursts is $|\gamip \ln \ep|$ in both
cases. This value depends on the spin of the star via $\ep$.  However,
the value of $\mdottmax$ at which the peak is reached is different. In
the case $\bet>0$, the maximum is reached at the pole, point \Cp{,}
and $\mdottmax = \bign \mip$, \eqr{equ:sum:mpmg}: almost constant
since $\bign$ is very close to $1$ for all the known bursters, unless
various effects (like the accretion processes discussed at the
beginning of this section) contribute to make $\bign$ a stronger
function of $\nu$. In the other case, $\bet < 0$, $\mdottmax = \bign
\mip \eps{\gamip}$. This value depends on the spin of the star more
strongly, especially if $\ep$ is given by what we called $\epm$: the
value due to mixing.  It is therefore easier to reconcile theory and
observations if $\bet<0$: the theoretical value for the quenching of
the bursts may be when the bursts are already too rare and dim to be
detectable above the fluctuating background accretion luminosity. The
case $\bet>0$ is still possible of course, but this would require a
very strong correction to our present understanding, since the value
predicted by theory would then correspond to $\mdottmax$ which is much
lower then current estimations. This latter case seems less likely.

We also mentioned cases where limits to $\xbase$ are set by
$\epsstari$, \eqr{equ:epsstar} and \appref{sec:lessone}. These
correspond to cases when $\rhoi \ge 1$ and $\delgami < 0$ (or when
$\rhoi <1$ and $\delgami >0$, see \appref{sec:lessone}). They
correspond to cases where the equator (or the pole) is always stable
for slow rotators. It is interesting to note that $\mdott$
corresponding to $\epsstari$, $\mdott = \bign \mip
{\epsstari}^{\gamip}$, is almost a constant. That is because of the
weak changes of $\bign$ and the fact that $\mip$, $\epsstari$ and
$\gamip$ are constants depending only on the burning physical
processes. This is a partial artefact of the cases we
treated. $\epsstari$ comes from equating $\mi \x{\gami} = \mip
\x{\gamip}$. The fact that $\nu$ and $\theta$ appear always together
with the same form in the $\xbase$'s (namely $\nu\sin\theta$) makes
the equality one equation in one unknown, the unknown being
$\nu\sin\theta$. Then, $\nu\sin\theta$ is fully determined and so are
the $\xbase$'s, which in turn make $\mdot$ fully determined and
constant. On the other hand, if the two $\xbase$'s on the two sides of
the equivalence were in fact different, most importantly
\emph{depending differently on $\nu$ and $\theta$}, something that
would happen for example if different mechanisms were at work or if
accretion physics were to change \eqr{equ:mdrel} and the relation
between $\mdotgen$ and $\mdot$, then equating the two boundaries for
regime $\ii$ would provide one equation in two unknowns and therefore
return a value for $\epsstari$ and the corresponding $\mdot$ that
would depend \emph{explicitly} on $\nu$. This would also result in
some $\mdottmax$ depend more strongly on $\nu$.

Very preliminary analysis of observational data shows that the
parameters $\bet$ and $\gam_*$ would need to be very large to account
for observations if only the effect of changing gravity is taken into
account. This can be seen from the fact that the span in $\mdot$
between the peak burst rate and the minimum is $\gamh\ln\ep$. Even
taking into account general relativistic effects $\ep$ is still very
close to $1$ even for fast rotators and that makes the logarithm
almost zero. On the other hand, when we considered mixing, we
suggested that $\xbasem$ would be mostly proportional to $\nu$, and so
would be $\epm$ (see \secref{sec:mixing}). That can lead to much
stronger effects. We do not study the exact form of $\xbasem$ in this
work, but deem it a very worthy direction of research.

Finally, we want to add that any other mechanism could explain the
decreasing burst rate if it would set a gap between the burst rate
from different ignition locations and it would provide a mean of
moving the ignition between these sites fast enough so that the
increase due to \inc{} would not be able to compensate for the initial
gap. We think that the local effective gravity and mixing are among
the most natural of such mechanisms.

\subsection{Future perspectives}

Ultimately, a joint effort of fitting to observations and running sets
of numerical models varying both $\mdotgen$ \emph{and} $\nu$ should
provide a test to the idea described above and, if successful, our
equations would provide the constraints that theory has to follow to
reproduce the observations. The $\theta$ dependent effects of mixing
should be included self consistently. We plan to perform a detailed
comparison with observations in the future, but here we mention some
considerations concerning the applications of \eqr{equ:buratebet} when
comparing to different sources. First of all the importance of
composition. We have seen that composition is important because it
determines the value of the coefficients $\Rbi$, $\mi$ and
$\mip$. \citet{art-2016-lampe-heg-gal} show that composition, even
small variations in metallicity, will also change the values of the
parameter $\alp$. It seems reasonable to expect that $\bet$ and $\gam$
may also be affected to some degree and therefore values may be
different from source to source. \citet{art-2007-piro-bild} and
\citet{art-2009-kee-lang-zand} also stress the effects of mixing.
\citet{art-2007-piro-bild} suggest that, to test the effects of spin
at a first order approximation, it would be sufficient to run
simulations varying the mass fraction of helium. The work of
\citet{art-2016-lampe-heg-gal} changed the composition (even if the
fraction of helium was tied to that of the metallicity of CNO species)
for a part of the range suggested by \citet{art-2007-piro-bild}. In
figure 1 of \citet{art-2016-lampe-heg-gal} it can be seen that indeed
the burst stabilized at appreciably different $\mdotgen$. Second, we
want to point out that following the evolution of a single source
different burning regimes will be experienced and therefore changes in
the parameters are to be expected when the burning regimes switch. For
example, the phase when burst rate decrease is observed could be
described by the mechanism we propose during burning of helium in a
mixed hydrogen--helium environment followed by the regime of delayed
mixed bursts described by \citet{art-2003-nara-heyl} and
\citet{art-2016-lampe-heg-gal}, where $\alp<0$ at the very latest
stages of accretion rate. This regime happens for a very narrow range
at high $\mdott$, before the bursts disappear. Even if this happens at
too high $\mdott$ to explain the observed decrease in burst rate in
all sources, it could still play a role in the very last stages of the
decreasing burst rate determined by the mechanism we propose. Another
example is what would happen when the switch is not between a burning
regime and burst stability, but between the burning regime $\ii$ and
the burning regime $\iip$. Locally the switch happens always when the
burst rate of regime $\iip$ becomes faster than the rate of regime
$\ii$. Suppose the switch from regime $\ii$ to $\iip$ takes place
initially at the equator, then the burst rate $\iip$ at the equator
can be either faster or slower than the burst rate $\ii$ at some other
colatitude. If it is higher, then the switch will take place also from
the point of view of the observer.  If it is slower, then the burst
rate will look like that of bursts of regime $\ii$ for higher
accretion rates until the burst rate of regime $\iip$ will eventually
overtake. In both cases the flame of the bursts will meet different
conditions across the surface of the star, giving for example
lightcurves with mixed properties: these effects would need to be
simulated with multi dimensional simulations (see below) and studying
a series of burning regimes would need the comparison of
\emph{compositions} of the diagrams of \figrefs{fig:TOS} -
\ref{fig:RBpos} to observations.

A striking feature of the relations of \secref{sec:ignitionmp} is how
the kind of the expected behaviour depends mostly on the sign of
$\delnn{\ii}{*}$ and the sign of the parameters $\gami$, $\gamip$.  In
case of switching of behaviour, controlled once again by $\gami$ and
$\gamip$, through $\delgami$, $\rhoi = \mi / \mip$ is also an
important parameter. The role of $\delnn{\ii}{*}$ is particularly
informative: since $\delnn{\ii}{*}$ combines both the contribution
from the bursting rate, via $\alpi$, $\beti$, and the boundaries in
mass accretion rate of a bursting regime, via $\gam_*$, one cannot say
that one of these aspects is much more important then the other. For
example, both a small, positive $\gamip$ or a strong, negative $\beti$
would give a negative $\deliip$. However, small or large is relative
to the value of $\alpi$. As a possibility, the values of $\gamip$ and
$\beti$ required to explain the observations could be determined,
respectively, by a weak dependence on $\xbase$ of the bound on
accretion rate or a strong positive dependence on $\xbase$ of $\yig$,
but there are of course other possible combinations. However, since
all these parameters, in one way or another, come from the ignition
depth of a specific burning regime (see \secref{sec:definitions}), and
this in turn depends quite strongly on the energy release rate, this
brings further evidence in support of the need that nuclear research
has to focus on better understanding the reaction rates, which set the
temperature of the burning region \citep{art-2011-schatz}. Another
important point to clarify, which also influences the ignition depth,
is the origin and magnitude of extra heat sources due for example to
the accretion process or further reactions in deeper layers as
speculated for explaining superbursts or NS cooling behaviour. The
dependence of these factors on the local conditions at different
colatitudes is key.

In an idealized procedure, one would have to input physical parameters
in 1D simulations and then extract the exponents $\alpi$, $\beti$,
$\gam_*$ and the masses $\mdotgen_*$ from the simulations changing the
effective gravity $\gef{}$, the local accretion rate $\mdotgen$ and
also the composition (see for example \citealt{art-2016-lampe-heg-gal}
and also \citealt{art-2017-gall-good-keek}), compare the results to
the constraints obtained from the data \citep[similar to those of][for
  example]{art-2003-corne-etal}, introduce more refined physical
processes and then repeat the procedure till convergence. The value of
$\alp$ for helium burning from the literature and simulations is well
established around $1 - 1.2$ \citep[see e.g.][]{rev-1998-bild,
  art-2016-lampe-heg-gal}, but the dependence over the other factors
should be explored, since information on the values of $\bet$ and
$\gam_*$ is scarce and should be measured more accurately. Note that
the results from 1D simulations would return the values $\alp$,
$\bett=\bet + \alp$ and $\gamt = \gam - 1$ in the case of the
dependence over gravity, while in the case of the dependence over
mixing the exponents would be directly $\alp$, $\bet$ and
$\gam_*$. However, one very important point to determine first is the
\emph{form} of the function $\xbasem$ (see \secrefs{sec:definitions}
and \ref{sec:mixing}).

This kind of fitting could benefit further from the use of global
multi-D simulations similar to those of \citet{art-2013-cavecchi-etal,
  art-2015-cavecchi-etal, art-2016-cavecchi-etal}. Application of the
analytical equations determines the most likely ignition colatitude
for the burst and indicates whether some part of the star may be
covered in partly exhaust fuel. This information could be used in the
global simulations in order to simulate flame spreading from the most
plausible colatitude with a reasonable surface distribution of the
fuel. The results could be compared to observed lightcurves in order
to test the agreement of additional details like for example burst
oscillations, or the exact profile of the lightcurves of the bursts
\citep[see e.g. the discussion in][]{art-2007-heg-cumm-gal-woos}.
Multi dimensional simulations would also be key in understanding the
differences in local initial conditions, or $\xbase$ in our
jargon. The advantage introduced by the analytical relations of this
paper is to make the comparison between observations and theoretical
models faster and possibly even indicate the direction in which to
search for refinements in order to match the observational
criteria. We have already pointed out one: $\alpi + \beti / \gamip <
0$.

Future large-area X-ray telescopes, such as the proposed Enhanced
X-ray Timing and Polarimetry mission
\citep[\textit{eXTP}][]{art-2016-zhang-etal} and the NASA Probe-class
mission concept \textit{STROBE-X} \citep{art-2017-wilhod-etal}, will
have improved sensitivity, all-sky monitoring and spectral-timing
capability. Analytical relations of the type given in this paper will
be particularly useful to interpret the high quality sequences of
burst and burst oscillation data expected from such missions, to
understand the details of burning and accretion physics.  Combining
that with multi-D simulations will allow a faster and more powerful
application of the phenomena associated with thermonuclear explosions
to the study of the properties of the underlying neutron stars, like
for example the use of type I bursts to tackle the problem of the
equation of state of the neutron star cores
\citep{rev-2013-miller-arxivpaper,rev-2016-watts-etal}.

\emph{Acknowledgements: } We thank L. Keek for comments on an earlier
version of the draft that improved its clarity. YC wishes to thank
P. Crumley for asking the right questions. YC is supported by the
European Union Horizon 2020 research and innovation programme under
the Marie Sklodowska-Curie Global Fellowship grant agreement No
703916. AW acknowledges support from ERC Starting Grant No. 639217
CSINEUTRONSTAR.

\appendix

\section{What if $\rhoi < 1$?}
\label{sec:lessone}

\figlessone{}

We stated at the beginning of \secref{sec:ignitionmp} the
condition $\rhoi \ge 1$, advocating the reason that $\rhoi < 1$
seems highly unlikely. We can explore quickly this alternative
here.

If $\rhoi < 1$ then the only possibility in order to have an
existing window for the bursts is $\delgami > 0$\footnote{Consider
  \eqrs{equ:mdotlimbad} and \eqref{equ:mdotlimbet}: it is needed
  that $\ml \x{\gaml} \le \mh \x{\gamh} \iff \ln(\ml \x{\gaml})
  \le \ln (\mh \x{\gamh}) \iff 0 \le \ln\rhoi - \delgami
  \ln\xbase$. Since $\ln\rhoi < 0$ and $\ln\xbase \le 0$, the
  inequality is satisfied only if $\delgami > 0$.}. Then, we can
see that \eqr{equ:burnexists} constrains $\xbase$ to be $\xbase
\le \xsi$, with $\xsi$ defined as in \eqr{equ:epsstar}. Since $\ep
\le \xbase$, it follows that $\ep \le \xsi$, or, equivalently, we
need $\thesi$, defined as in \eqr{equ:thesi}, to exist. In summary
\begin{equation}
  \label{equ:exist3}
  \ep  \le \xbase \le \xsi \;\; if \;\;
  \rhoi < 1 \anda \delgami > 0 \anda \ep \le \epsstari
\end{equation}
The first conclusion is that when $\rhoi < 1$ ignition at the pole is
forbidden. Second, the requirement $\ep \le \xsi$ implies that very
slow rotators would not show bursts.

We will proceed to study the case $\alpi > 0$. The procedure follows
very closely the one of \secref{sec:ignitionmp}, with the only
difference set by the fact that the upper limit is $\xsi$ and not
$1$. If we study the burst rate evolution for a single source as a
function of $\mdot$, we obtain \figref{fig:lessone}. The results when
$\beti < 0$ are identical to the upper panel of \figref{fig:RBneg},
apart from the fact that the last case is impossible. When $\beti > 0$
the results resemble those of the upper panel of \figref{fig:RBpos},
with the loss of the segments corresponding to ignition at the pole
and the absence of the last case. Of course, if $\beti = 0$, the rate
grows always $\propto \mdot^{\alpi}$.

There is an extra interesting detail. If we think that $\gamip>0$,
the maximum $\mdot$ at which bursts will take place is less than
the theoretical local one, always, even if detectability were not
an issue. That is because
\begin{equation}
  \mdotgen_{\rm{p, D} \equiv \rm{C}} =  \mip \, {\xsi}^{\gamip} = 
  \mip \, \rhoi^{\gamip/\delgami}
  < \mip
\end{equation}

In \secref{sec:exabild} we noted how the numerical factor for
$\mdotgen_{2, \hig}$, the upper bound for the helium ignition in a
mixed hydrogen--helium environment, \eqr{equ:mdotHHEhigh}, is much
smaller than the one for $\mdotgen_{2, \low}$, the lower bound,
\eqr{equ:mdotHEhigh}. That led us to explore the possibility
$\rhoi < 1$. However, we still deem this possibility less likely
than the one treated in the main text.

\section{Extra cases: $\alpi < 0$ and $\alpi = 0$.}
\label{sec:extra}

Here we report the extra, less physical cases, $\alpi < 0$ and $\alpi
= 0$, mainly for mathematical completeness.

\subsection{Case $\alpi < 0$}
\label{sec:alphaneg}

Both \citet{art-2003-nara-heyl} with linearized calculations and
\citet{art-2016-lampe-heg-gal} with \KEP simulations found some cases
where $\alpi < 0$. The window of accretion rate where that happens is
relatively small; however, that induces us to discuss this case. The
evolution of the bursting rate as a function of $\mdot$ for a single
source is shown in \figrefs{fig:negRBneg} ($\beti<0$) and
\ref{fig:negRBpos} ($\beti>0$). In this case it is known that $\deliip
/ \gamip < \alpi < 0$ when we need it, while the sign of
$\delii/\gami$ is not known for the cases needed: it depends on the
sign of $-\delii$ when $\beti <0$ (\figref{fig:negRBneg}) and on the
sign of $\delii$ when $\beti>0$ (\figref{fig:negRBpos}). When the
$\beti=0$ the rate always goes $\propto \mdot^{\alpi}$: decreasing
since $\alpi<0$.

\fignegapp{}

\subsection{Case $\alpi = 0$}
\label{sec:alphazer}

Based on \eqr{equ:buratebet}, we know that $\R = \Rbi \x{\beti}$ for
all available ignition colatitudes. This is the most unnatural case,
since the burning rate does not depend on $\mdot$, but it changes of
course depending on the colatitude. The evolution of the bursting rate
as a function of $\mdot$ for a single source is shown in
\figrefs{fig:zerRBneg} ($\beti<0$) and \ref{fig:zerRBpos}
($\beti>0$). Both the signs of $\deliip/\gamip = \beti/\gamip$ and
$\delii/\gami = \beti/\gami$ are known when needed. When $\beti=0$ the
rate is constant.

\figzerapp{}

$\phantom{a}$
\newpage
\biba

\label{lastpage}


\begin{thebibliography}{}
\expandafter\ifx\csname natexlab\endcsname\relax\def\natexlab#1{#1}\fi
\providecommand{\url}[1]{\href{#1}{#1}}

\bibitem[{{AlGendy} \& {Morsink}(2014)}]{art-2014-alge-mor}
{AlGendy}, M., \& {Morsink}, S.~M. 2014, \apj, 791, 78

\bibitem[{{Altamirano} {et~al.}(2010){Altamirano}, {Watts}, {Kalamkar},
  {Homan}, {Yang}, {Casella}, {Linares}, {Patruno}, {Armas Padilla},
  {Cavecchi}, {Degenaar}, {Kaur}, {van der Klis}, {Rea}, \&
  {Wijnands}}]{atel-2010-alta-etal}
{Altamirano}, D., {Watts}, A., {Kalamkar}, M., {et~al.} 2010, The Astronomer's
  Telegram, 2932, 1

\bibitem[{{Bildsten}(1998)}]{rev-1998-bild}
{Bildsten}, L. 1998, in NATO ASIC Proc. 515: The Many Faces of Neutron Stars,
  ed. R.~{Buccheri}, J.~{van Paradijs}, \& A.~{Alpar}, 419--+

\bibitem[{{Bildsten}(2000)}]{rev-2000-bild}
{Bildsten}, L. 2000, in American Institute of Physics Conference Series, Vol.
  522, American Institute of Physics Conference Series, ed. S.~S. {Holt} \&
  W.~W. {Zhang}, 359--369

\bibitem[{{Bildsten} \& {Brown}(1997)}]{art-1997-bild-brown}
{Bildsten}, L., \& {Brown}, E.~F. 1997, \apj, 477, 897

\bibitem[{{Brown} {et~al.}(1998){Brown}, {Bildsten}, \&
  {Rutledge}}]{art-1998-brown-bild-rut}
{Brown}, E.~F., {Bildsten}, L., \& {Rutledge}, R.~E. 1998, \apjl, 504, L95

\bibitem[{{Brown} \& {Cumming}(2009)}]{art-2009-brow-cum}
{Brown}, E.~F., \& {Cumming}, A. 2009, \apj, 698, 1020

\bibitem[{{Cavecchi} {et~al.}(2016){Cavecchi}, {Levin}, {Watts}, \&
  {Braithwaite}}]{art-2016-cavecchi-etal}
{Cavecchi}, Y., {Levin}, Y., {Watts}, A.~L., \& {Braithwaite}, J. 2016, \mnras,
  459, 1259

\bibitem[{{Cavecchi} {et~al.}(2013){Cavecchi}, {Watts}, {Braithwaite}, \&
  {Levin}}]{art-2013-cavecchi-etal}
{Cavecchi}, Y., {Watts}, A.~L., {Braithwaite}, J., \& {Levin}, Y. 2013, \mnras,
  434, 3526

\bibitem[{{Cavecchi} {et~al.}(2015){Cavecchi}, {Watts}, {Levin}, \&
  {Braithwaite}}]{art-2015-cavecchi-etal}
{Cavecchi}, Y., {Watts}, A.~L., {Levin}, Y., \& {Braithwaite}, J. 2015, \mnras,
  448, 445

\bibitem[{{Cooper} \& {Narayan}(2006{\natexlab{a}})}]{art-2006-coop-nara-a}
{Cooper}, R.~L., \& {Narayan}, R. 2006{\natexlab{a}}, \apjl, 648, L123

\bibitem[{{Cooper} \& {Narayan}(2006{\natexlab{b}})}]{art-2006-coop-nara-b}
---. 2006{\natexlab{b}}, \apj, 652, 584

\bibitem[{{Cooper} \& {Narayan}(2007{\natexlab{a}})}]{art-2007-coop-nara-a}
---. 2007{\natexlab{a}}, \apjl, 657, L29

\bibitem[{{Cooper} \& {Narayan}(2007{\natexlab{b}})}]{art-2007-coop-nara-b}
---. 2007{\natexlab{b}}, \apj, 661, 468

\bibitem[{{Cornelisse} {et~al.}(2003){Cornelisse}, {in't Zand}, {Verbunt},
  {Kuulkers}, {Heise}, {den Hartog}, {Cocchi}, {Natalucci}, {Bazzano}, \&
  {Ubertini}}]{art-2003-corne-etal}
{Cornelisse}, R., {in't Zand}, J.~J.~M., {Verbunt}, F., {et~al.} 2003, \aap,
  405, 1033

\bibitem[{{Cumming}(2003)}]{art-2003-cum}
{Cumming}, A. 2003, \apj, 595, 1077

\bibitem[{{Cumming}(2004)}]{art-2004-cum}
---. 2004, Nuclear Physics B Proceedings Supplements, 132, 435

\bibitem[{{Cumming} \& {Bildsten}(2000)}]{art-2000-cum-bild}
{Cumming}, A., \& {Bildsten}, L. 2000, \apj, 544, 453

\bibitem[{{Cyburt} {et~al.}(2016){Cyburt}, {Amthor}, {Heger}, {Johnson},
  {Keek}, {Meisel}, {Schatz}, \& {Smith}}]{art-2016-cyb-etal}
{Cyburt}, R.~H., {Amthor}, A.~M., {Heger}, A., {et~al.} 2016, \apj, 830, 55

\bibitem[{{Cyburt} {et~al.}(2010){Cyburt}, {Amthor}, {Ferguson}, {Meisel},
  {Smith}, {Warren}, {Heger}, {Hoffman}, {Rauscher}, {Sakharuk}, {Schatz},
  {Thielemann}, \& {Wiescher}}]{art-2010-cyb-etal}
{Cyburt}, R.~H., {Amthor}, A.~M., {Ferguson}, R., {et~al.} 2010, \apjs, 189,
  240

\bibitem[{{Davids} {et~al.}(2011){Davids}, {Cyburt}, {Jos{\'e}}, \&
  {Mythili}}]{art-2011-dav-cyb-etal}
{Davids}, B., {Cyburt}, R.~H., {Jos{\'e}}, J., \& {Mythili}, S. 2011, \apj,
  735, 40

\bibitem[{{Fisker} {et~al.}(2007){Fisker}, {Tan}, {G{\"o}rres}, {Wiescher}, \&
  {Cooper}}]{art-2007-fisk-etal}
{Fisker}, J.~L., {Tan}, W., {G{\"o}rres}, J., {Wiescher}, M., \& {Cooper},
  R.~L. 2007, \apj, 665, 637

\bibitem[{{Fujimoto}(1993)}]{art-1993-fuji}
{Fujimoto}, M.~Y. 1993, \apj, 419, 768

\bibitem[{{Fujimoto} {et~al.}(1981){Fujimoto}, {Hanawa}, \&
  {Miyaji}}]{art-1981-fuji-han-miy}
{Fujimoto}, M.~Y., {Hanawa}, T., \& {Miyaji}, S. 1981, \apj, 247, 267

\bibitem[{{Galloway} {et~al.}(2010){Galloway}, {in't Zand}, {Chenevez}, {Keek},
  \& {Brandt}}]{rev-2010-gal-etal}
{Galloway}, D., {in't Zand}, J., {Chenevez}, J., {Keek}, L., \& {Brandt}, S.
  2010, in COSPAR Meeting, Vol.~38, 38th COSPAR Scientific Assembly, 2445

\bibitem[{{Galloway} {et~al.}(2017){Galloway}, {Goodwin}, \&
  {Keek}}]{art-2017-gall-good-keek}
{Galloway}, D.~K., {Goodwin}, A.~J., \& {Keek}, L. 2017, \pasa, 34, e019

\bibitem[{{Hanawa} \& {Fujimoto}(1984)}]{art-1984-hana-fuji}
{Hanawa}, T., \& {Fujimoto}, M.~Y. 1984, \pasj, 36, 199

\bibitem[{{Hartman} {et~al.}(2003){Hartman}, {Chakrabarty}, {Galloway}, {Muno},
  {Savov}, {Mendez}, {van Straaten}, \& {Di Salvo}}]{art-2003-hart-etal}
{Hartman}, J.~M., {Chakrabarty}, D., {Galloway}, D.~K., {et~al.} 2003, in
  Bulletin of the American Astronomical Society, Vol.~35, AAS/High Energy
  Astrophysics Division \#7, 865

\bibitem[{{Heger} {et~al.}(2007{\natexlab{a}}){Heger}, {Cumming}, {Galloway},
  \& {Woosley}}]{art-2007-heg-cumm-gal-woos}
{Heger}, A., {Cumming}, A., {Galloway}, D.~K., \& {Woosley}, S.~E.
  2007{\natexlab{a}}, \apjl, 671, L141

\bibitem[{{Heger} {et~al.}(2007{\natexlab{b}}){Heger}, {Cumming}, \&
  {Woosley}}]{art-2007-heg-cum-woos}
{Heger}, A., {Cumming}, A., \& {Woosley}, S.~E. 2007{\natexlab{b}}, \apj, 665,
  1311

\bibitem[{{Hessels} {et~al.}(2006){Hessels}, {Ransom}, {Stairs}, {Freire},
  {Kaspi}, \& {Camilo}}]{art-2006-hessels-j-etal}
{Hessels}, J.~W.~T., {Ransom}, S.~M., {Stairs}, I.~H., {et~al.} 2006, Science,
  311, 1901

\bibitem[{{Inogamov} \& {Sunyaev}(1999)}]{art-1999-ino-suny}
{Inogamov}, N.~A., \& {Sunyaev}, R.~A. 1999, Astronomy Letters, 25, 269

\bibitem[{{Inogamov} \& {Sunyaev}(2010)}]{art-2010-inoga-suny}
---. 2010, Astronomy Letters, 36, 848

\bibitem[{{Kajava} {et~al.}(2014){Kajava}, {N{\"a}ttil{\"a}}, {Latvala},
  {Pursiainen}, {Poutanen}, {Suleimanov}, {Revnivtsev}, {Kuulkers}, \&
  {Galloway}}]{art-2014-kaja-etal}
{Kajava}, J.~J.~E., {N{\"a}ttil{\"a}}, J., {Latvala}, O.-M., {et~al.} 2014,
  \mnras, 445, 4218

\bibitem[{{Keek} {et~al.}(2014){Keek}, {Cyburt}, \&
  {Heger}}]{art-2014-keek-cyb-heger}
{Keek}, L., {Cyburt}, R.~H., \& {Heger}, A. 2014, \apj, 787, 101

\bibitem[{{Keek} {et~al.}(2009){Keek}, {Langer}, \& {in't
  Zand}}]{art-2009-kee-lang-zand}
{Keek}, L., {Langer}, N., \& {in't Zand}, J.~J.~M. 2009, \aap, 502, 871

\bibitem[{{Lampe} {et~al.}(2016){Lampe}, {Heger}, \&
  {Galloway}}]{art-2016-lampe-heg-gal}
{Lampe}, N., {Heger}, A., \& {Galloway}, D.~K. 2016, \apj, 819, 46

\bibitem[{{Linares} {et~al.}(2012){Linares}, {Altamirano}, {Chakrabarty},
  {Cumming}, \& {Keek}}]{art-2012-lin-etal}
{Linares}, M., {Altamirano}, D., {Chakrabarty}, D., {Cumming}, A., \& {Keek},
  L. 2012, \apj, 748, 82

\bibitem[{{Malone} {et~al.}(2011){Malone}, {Nonaka}, {Almgren}, {Bell}, \&
  {Zingale}}]{art-2011-malo-etal}
{Malone}, C.~M., {Nonaka}, A., {Almgren}, A.~S., {Bell}, J.~B., \& {Zingale},
  M. 2011, \apj, 728, 118

\bibitem[{{Miller}(2013)}]{rev-2013-miller-arxivpaper}
{Miller}, M.~C. 2013, ArXiv e-prints, arXiv:1312.0029

\bibitem[{{Muno} {et~al.}(2002){Muno}, {Chakrabarty}, {Galloway}, \&
  {Psaltis}}]{art-2002-muno-etal}
{Muno}, M.~P., {Chakrabarty}, D., {Galloway}, D.~K., \& {Psaltis}, D. 2002,
  \apj, 580, 1048

\bibitem[{{Narayan} \& {Heyl}(2003)}]{art-2003-nara-heyl}
{Narayan}, R., \& {Heyl}, J.~S. 2003, \apj, 599, 419

\bibitem[{{Philippov} {et~al.}(2016){Philippov}, {Rafikov}, \&
  {Stone}}]{art-2016-philippov-rafi-stone}
{Philippov}, A.~A., {Rafikov}, R.~R., \& {Stone}, J.~M. 2016, \apj, 817, 62

\bibitem[{{Piro} \& {Bildsten}(2007)}]{art-2007-piro-bild}
{Piro}, A.~L., \& {Bildsten}, L. 2007, \apj, 663, 1252

\bibitem[{{Schatz}(2011)}]{art-2011-schatz}
{Schatz}, H. 2011, Progress in Particle and Nuclear Physics, 66, 277

\bibitem[{{Schatz} {et~al.}(2001){Schatz}, {Aprahamian}, {Barnard}, {Bildsten},
  {Cumming}, {Ouellette}, {Rauscher}, {Thielemann}, \&
  {Wiescher}}]{art-2001-schatz-etal}
{Schatz}, H., {Aprahamian}, A., {Barnard}, V., {et~al.} 2001, Physical Review
  Letters, 86, 3471

\bibitem[{{Schatz} {et~al.}(2014){Schatz}, {Gupta}, {M{\"o}ller}, {Beard},
  {Brown}, {Deibel}, {Gasques}, {Hix}, {Keek}, {Lau}, {Steiner}, \&
  {Wiescher}}]{art-2014-schatz-etal}
{Schatz}, H., {Gupta}, S., {M{\"o}ller}, P., {et~al.} 2014, \nat, 505, 62

\bibitem[{{Spruit}(1999)}]{art-1999-spruit}
{Spruit}, H.~C. 1999, \aap, 349, 189

\bibitem[{{Spruit}(2002)}]{art-2002-spruit}
---. 2002, \aap, 381, 923

\bibitem[{{Strohmayer} \& {Bildsten}(2006)}]{rev-2003-2006-stro-bild-book}
{Strohmayer}, T., \& {Bildsten}, L. 2006, {New views of thermonuclear bursts},
  ed. W.~H.~G. {Lewin} \& M.~{van der Klis} (Cambridge University Press),
  113--156

\bibitem[{{Tan} {et~al.}(2007){Tan}, {Fisker}, {G{\"o}rres}, {Couder}, \&
  {Wiescher}}]{art-2007-tan-etal}
{Tan}, W.~P., {Fisker}, J.~L., {G{\"o}rres}, J., {Couder}, M., \& {Wiescher},
  M. 2007, Physical Review Letters, 98, 242503

\bibitem[{{van Paradijs} {et~al.}(1988){van Paradijs}, {Penninx}, \&
  {Lewin}}]{art-1988-par-pen-lew}
{van Paradijs}, J., {Penninx}, W., \& {Lewin}, W.~H.~G. 1988, \mnras, 233, 437

\bibitem[{{Watts}(2012)}]{rev-2012-watts}
{Watts}, A.~L. 2012, \araa, 50, 609

\bibitem[{{Watts} {et~al.}(2016){Watts}, {Andersson}, {Chakrabarty}, {Feroci},
  {Hebeler}, {Israel}, {Lamb}, {Miller}, {Morsink}, {{\"O}zel}, {Patruno},
  {Poutanen}, {Psaltis}, {Schwenk}, {Steiner}, {Stella}, {Tolos}, \& {van der
  Klis}}]{rev-2016-watts-etal}
{Watts}, A.~L., {Andersson}, N., {Chakrabarty}, D., {et~al.} 2016, Reviews of
  Modern Physics, 88, 021001

\bibitem[{{Wijnands} {et~al.}(2013){Wijnands}, {Degenaar}, \&
  {Page}}]{art-2013-wij-dege-page}
{Wijnands}, R., {Degenaar}, N., \& {Page}, D. 2013, \mnras, 432, 2366

\bibitem[{{Wilson-Hodge} {et~al.}(2017){Wilson-Hodge}, {Ray}, {Gendreau},
  {Chakrabarty}, {Feroci}, {Maccarone}, {Arzoumanian}, {Remillard}, {Wood},
  {Griffith}, \& {STROBE-X Collaboration}}]{art-2017-wilhod-etal}
{Wilson-Hodge}, C.~A., {Ray}, P.~S., {Gendreau}, K., {et~al.} 2017, in American
  Astronomical Society Meeting Abstracts, Vol. 229, American Astronomical
  Society Meeting Abstracts, 309.04

\bibitem[{{Woosley} {et~al.}(2004){Woosley}, {Heger}, {Cumming}, {Hoffman},
  {Pruet}, {Rauscher}, {Fisker}, {Schatz}, {Brown}, \&
  {Wiescher}}]{art-2004-woos-etal}
{Woosley}, S.~E., {Heger}, A., {Cumming}, A., {et~al.} 2004, \apjs, 151, 75

\bibitem[{{Zamfir} {et~al.}(2014){Zamfir}, {Cumming}, \&
  {Niquette}}]{art-2014-zam-cumm-niq}
{Zamfir}, M., {Cumming}, A., \& {Niquette}, C. 2014, \mnras, 445, 3278

\bibitem[{{Zhang} {et~al.}(2016){Zhang}, {Feroci}, {Santangelo}, {Dong},
  {Feng}, {Lu}, {Nandra}, {Wang}, {Zhang}, {Bozzo}, {Brandt}, {De Rosa}, {Gou},
  {Hernanz}, {van der Klis}, {Li}, {Liu}, {Orleanski}, {Pareschi}, {Pohl},
  {Poutanen}, {Qu}, {Schanne}, {Stella}, {Uttley}, {Watts}, {Xu}, {Yu}, {in 't
  Zand}, {Zane}, {Alvarez}, {Amati}, {Baldini}, {Bambi}, {Basso},
  {Bhattacharyya S.}, {}, {Belloni}, {Bellutti}, {Bianchi}, {Brez}, {Bursa},
  {Burwitz}, {Budtz-J{\o}rgensen}, {Caiazzo}, {Campana}, {Cao}, {Casella},
  {Chen}, {Chen}, {Chen}, {Chen}, {Chen}, {Chen}, {Civitani}, {Coti Zelati},
  {Cui}, {Cui}, {Dai}, {Del Monte}, {de Martino}, {Di Cosimo}, {Diebold},
  {Dovciak}, {Donnarumma}, {Doroshenko}, {Esposito}, {Evangelista}, {Favre},
  {Friedrich}, {Fuschino}, {Galvez}, {Gao}, {Ge}, {Gevin}, {Goetz}, {Han},
  {Heyl}, {Horak}, {Hu}, {Huang}, {Huang}, {Hudec}, {Huppenkothen}, {Israel},
  {Ingram}, {Karas}, {Karelin}, {Jenke}, {Ji}, {Korpela}, {Kunneriath},
  {Labanti}, {Li}, {Li}, {Li}, {Liang}, {Limousin}, {Lin}, {Ling}, {Liu},
  {Liu}, {Liu}, {Lu}, {Lund}, {Lai}, {Luo}, {Luo}, {Ma}, {Mahmoodifar},
  {Marisaldi}, {Martindale}, {Meidinger}, {Men}, {Michalska}, {Mignani},
  {Minuti}, {Motta}, {Muleri}, {Neilsen}, {Orlandini}, {Pan}, {Patruno},
  {Perinati}, {Picciotto}, {Piemonte}, {Pinchera}, {Rachevski A.}, {Rapisarda},
  {Rea}, {Rossi}, {Rubini}, {Sala}, {Shu}, {Sgro}, {Shen}, {Soffitta}, {Song},
  {Spandre}, {Stratta}, {Strohmayer}, {Sun}, {Svoboda}, {Tagliaferri},
  {Tenzer}, {Hong}, {Taverna}, {Torok}, {Turolla}, {Vacchi}, {Wang}, {Walton},
  {Wang}, {Wang}, {Wang}, {Wang}, {Weng}, {Wilms}, {Winter}, {Wu}, {Wu},
  {Xiong}, {Xu}, {Xue}, {Yan}, {Yang}, {Yang}, {Yang}, {Yuan}, {Yuan}, {Yuan},
  {Zampa}, {Zampa}, {Zdziarski}, {Zhang}, {Zhang}, {Zhang}, {Zhang}, {Zhang},
  {Zhang}, {Zheng}, {Zhou}, \& {Zhou X.~L.}}]{art-2016-zhang-etal}
{Zhang}, S.~N., {Feroci}, M., {Santangelo}, A., {et~al.} 2016, in \procspie,
  Vol. 9905, Space Telescopes and Instrumentation 2016: Ultraviolet to Gamma
  Ray, 99051Q

\bibitem[{{Zingale} {et~al.}(2015){Zingale}, {Malone}, {Nonaka}, {Almgren}, \&
  {Bell}}]{art-2015-zing-etal}
{Zingale}, M., {Malone}, C.~M., {Nonaka}, A., {Almgren}, A.~S., \& {Bell},
  J.~B. 2015, \apj, 807, 60

\bibitem[{{Zingale} {et~al.}(2001){Zingale}, {Timmes}, {Fryxell}, {Lamb},
  {Olson}, {Calder}, {Dursi}, {Ricker}, {Rosner}, {MacNeice}, \&
  {Tufo}}]{art-2001-zing-etal}
{Zingale}, M., {Timmes}, F.~X., {Fryxell}, B., {et~al.} 2001, \apjs, 133, 195

\end{thebibliography}
\end{document}